\documentclass[aps,prl,twocolumn,superscriptaddress,10pt,english]{revtex4-1}
\usepackage[T1]{fontenc}
\usepackage[utf8]{inputenc}
\usepackage{xcolor}
\usepackage{babel}
\usepackage{graphicx}
\usepackage[hidelinks]{hyperref}
\newcommand{\dd}{\mathrm{d}}

\usepackage{babel}

\begin{document}

\title{Gravitational Waves from Holographic Neutron Star Mergers}

\author{Christian Ecker} 
\email{c.ecker@uu.nl} 
\affiliation{Institute for Theoretical Physics and Center for Extreme Matter and Emergent Phenomena, Utrecht University, Leuvenlaan 4, 3584 CE Utrecht, The Netherlands}
\affiliation{Instituut-Lorentz, $\Delta$ITP, Universiteit Leiden, P.O. Box 9506, 2300 RA Leiden, The Netherlands}
\author{Matti J\"arvinen} 
\email{m.o.jarvinen@uu.nl}
\affiliation{Institute for Theoretical Physics and Center for Extreme Matter and Emergent Phenomena, Utrecht University, Leuvenlaan 4, 3584 CE Utrecht, The Netherlands}
\affiliation{Department of Physics and Helsinki Institute of Physics, P.O. Box 64, FI-00014 University of Helsinki, Finland}
\author{Govert Nijs} 
\email{g.h.nijs@uu.nl} 
\affiliation{Institute for Theoretical Physics and Center for Extreme Matter and Emergent Phenomena, Utrecht University, Leuvenlaan 4, 3584 CE Utrecht, The Netherlands}
\author{Wilke van der Schee} 
\email{w.vanderschee@uu.nl} 
\affiliation{Institute for Theoretical Physics and Center for Extreme Matter and Emergent Phenomena, Utrecht University, Leuvenlaan 4, 3584 CE Utrecht, The Netherlands}

\begin{abstract}
We simulate the merger of binary neutron stars and analyze the spectral properties of their gravitational waveforms.
For the stars we construct hybrid equations of state (EoSs) with a standard nuclear matter EoS at low densities, transitioning to a state-of-the-art holographic EoS in the otherwise intractable high density regime. 
Depending on the transition density the characteristic frequencies in the spectrum 
produced from the hybrid EoSs are shifted to significantly lower values as compared to the pure nuclear matter EoS.
The highest rest-mass density reached outside a possible black hole horizon is approximately $1.1 \cdot 10^{15}$ g/cm$^3$, 
which for the holographic model is below the density of the deconfined quark matter phase.

\end{abstract}

\maketitle

\noindent
\textbf{1. Introduction.}
Gravitational waves (GWs) provide us a direct view on the most violent events in our universe \cite{Abbott:2016blz,TheLIGOScientific:2017qsa}, among which there will be tens of neutron star mergers in the near future. While General Relativity (GR) is by now a well understood theory, the merger of two neutron stars is clouded with considerably more uncertainty (see  \cite{Baiotti:2016qnr} for an excellent review). One profound reason is that the fundamental theory of quarks and gluons (QCD) is strongly coupled, and that the dense state of neutron stars is inaccessible to the lattice. In this Letter we hence  use a strongly coupled (holographic) model and study its implications on the power spectral density (PSD) of GWs after a neutron star merger.

Our current knowledge about the equation of state (EoS) in neutron stars comes from the low-density region (with energy density $\lesssim 0.2 \text{ GeV/fm}^3$) where it is constrained by nuclear physics \cite{Gandolfi:2011xu} and at very high densities ($\gtrsim 10 \text{ GeV/fm}^3$), where QCD is perturbative \cite{Gorda:2018gpy}. At intermediate densities not much is certain, but the EoS can be constrained by limiting the speed of sound as well as observations of mass-radius relationships and recently by the bound on the tidal deformability coming from the GW170817 merger measured by LIGO/Virgo \cite{TheLIGOScientific:2017qsa}. Applying these constraints it is likely that a phase transition is present \cite{Annala:2019puf}, which may indicate the presence of quark matter at high densities (see also \cite{Annala:2017llu,Most:2018hfd,Most:2018eaw, Bauswein:2018bma}).

Since QCD at intermediate densities is strongly coupled it is possible to obtain qualitative insights by employing holography. 
We use a holographic model to derive EoSs which are then used as input in simulations of equal-mass neutron star mergers performed with the \texttt{Einstein Toolkit} \cite{Loffler:2011ay}. 
Especially interesting are mergers of intermediately massive stars ($M\!\approx\!1.3-1.4 M_\odot$) because they can lead to the formation of a meta-stable hypermassive neutron star (HMNS), whose GW signal %
encodes characteristic information on the EoS \cite{Takami:2014zpa,Takami:2014tva,Radice:2016rys,Maione:2017aux}.

Finally we show how the PSD depends on both the parameters of our equation of state
and the neutron star mass,
and comment on how this compares with other EoSs. Future measurements by  advanced LIGO and the Einstein Telescope will be able to experimentally distinguish these different EoSs (see also \cite{Takami:2014zpa,Takami:2014tva} for earlier work with traditional nuclear matter EoSs).

\noindent
\textbf{2. Holographic Model.}
The holographic model (V-QCD), which we use to describe strongly interacting dense nuclear and deconfined quark matter~\cite{Jarvinen:2011qe}, is obtained through a fusion of two building blocks: %
improved holographic QCD~\cite{Gursoy:2007cb}, a bottom-up model for the gluon sector inspired by five-dimensional non-critical string theory, %
and a framework based on tachyonic D-brane actions for the flavor sector~\cite{Bigazzi:2005md,Casero:2007ae}. 
Details of the model are chosen to reproduce a number of features of QCD. This includes linear confinement, asymptotic freedom, chiral symmetry breaking, qualitatively reasonable hadron spectrum and a finite temperature phase diagram with structure as expected from QCD~\cite{Bigazzi:2005md,Casero:2007ae,Gursoy:2007cb,Jarvinen:2011qe,Alho:2012mh,Alho:2013hsa,Arean:2013tja,Jarvinen:2015ofa}.
The thermodynamics is tuned to match lattice data at temperatures right above the confinement/chiral transition~\cite{Gursoy:2009jd,Jokela:2018ers}.

This model has a nuclear phase \cite{Ishii:2019gta} and deconfined quark matter phase \cite{Jokela:2018ers,Chesler:2019osn,Hoyos:2016zke} at low and high densities respectively.
For the nuclear phase, essential for neutron stars, a simple method is used~\cite{Rozali:2007rx,Li:2015uea,Ishii:2019gta}, which approximates nuclear matter as a homogeneous field. 
The EoS for dense nuclear matter is stiff in this approach (i.e., the speed of sound is relatively high), and depends among other things on the coupling $b$ between the baryons and the chiral condensate (see \cite{Ishii:2019gta} for a complete exposition). 

\noindent
\textbf{3. Hybrid equations of state.}
\begin{table}[h!tb]
 \caption{
 Properties of irrotational neutron stars with hybrid SLy+VQCD and the SLy EoSs. Here
$\rho_\mathrm{NM}$ ($\rho_\mathrm{QM}$) is the nuclear (quark) matter density at the transition.
}

 \begin{tabular}{cc|cc|ccc}\hline\hline
  $\frac{\rho_m}{\rho_s}$ & $b$ &  $\frac{\rho_\mathrm{NM}}{\rho_s}$ &  $\frac{\rho_\mathrm{QM}}{\rho_s}$ & $\frac{M_\mathrm{max}}{M_\odot}$ & $R(1.4M_\odot)[km]$ & $\Lambda(1.4M_\odot)$ %
  \\
  \hline\hline
  1.44 & 10.45 & 4.29 & 7.93 & 2.25 & 13.0 & 680%
  \\
  1.61 & 10.50 & 4.83 & 8.57& 2.17 & 12.6 & 550 %
  \\
  1.77 & 10.55 & 5.37 & 9.24 & 2.10 & 12.3 & 470%
  \\
  1.94 & 10.60 & 5.86 & 9.79 & 2.04  & 12.1  & 410%
  \\
  2.10 & 10.65 & 6.23 & 10.21 & 2.00 & 12.0 & 370%
  \\
 \hline
 SLy & -- & -- & -- & 2.05 &  11.7 & 300\\
 \hline
\end{tabular}
\label{Tab:EOSs1} 
\end{table}
The weakly coupled, low density regime of cold nuclear matter is better described by using traditional effective methods. %
Therefore we construct ``hybrid'' equations of state composed of: i) 
SLy EoS~\cite{Haensel:1993zw,Douchin:2001sv} for low density nuclear matter; ii) V-QCD EoS for dense nuclear and quark matter described in Section 2. %
We choose the matching density $\rho_m$ between SLy and V-QCD to lie
at $1.5$ \textemdash $2$ times the nuclear saturation density $\rho_s\!=\!2.04 \cdot 10^{14}\,$g/cm$^{3}$ and then fix $b$ as well as the normalization of the baryon action $c_b$ by requiring that the pressure and its derivative with respect to the chemical potential are continuous at the matching point. 

\begin{figure}[htb]
\center
 \includegraphics[width=0.85\linewidth]{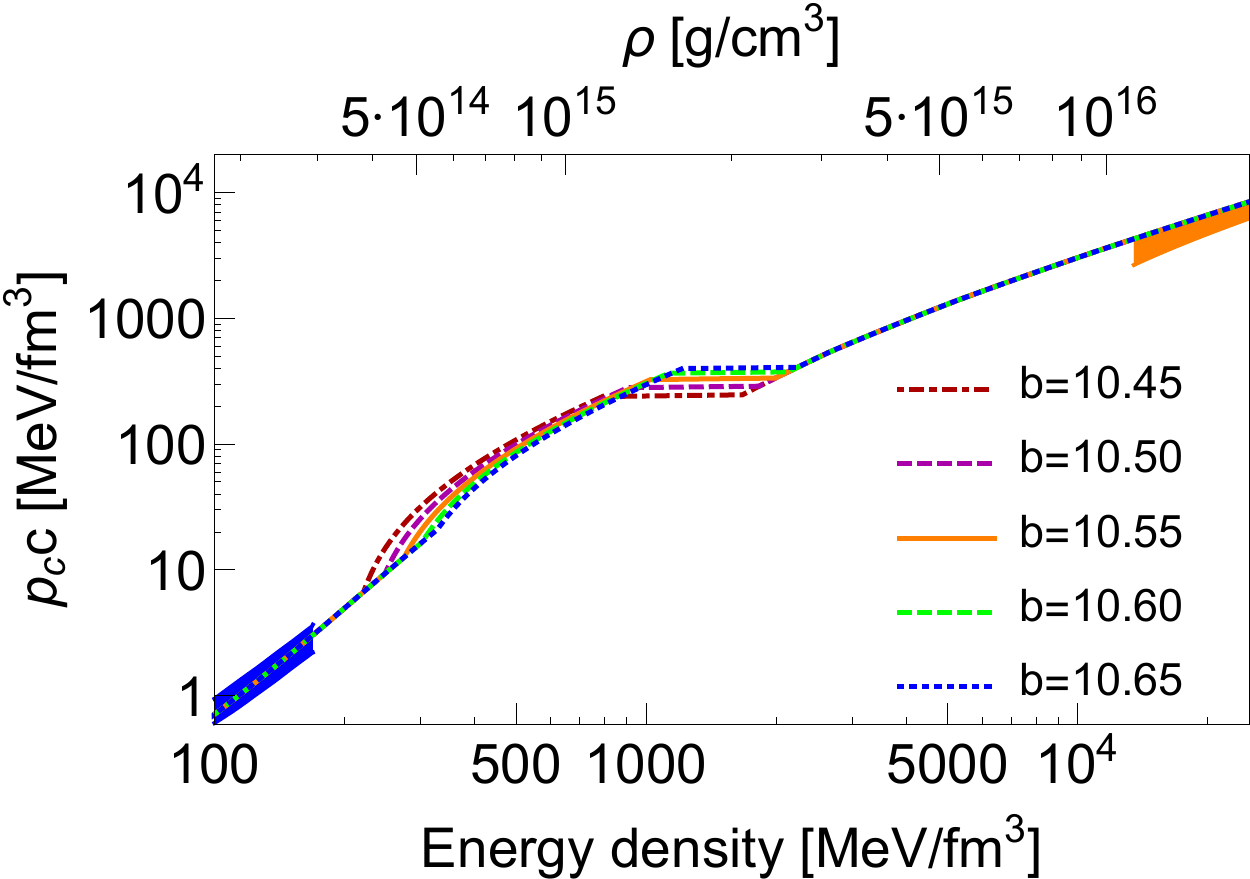}
\caption{The constructed family of the hybrid EoSs. %
Blue and orange regions show the error bands for QCD from effective Lagrangians and perturbative QCD, respectively.
The $b$-dependence on the $\rho$-axis (which is a non-trivial function of the energy density) is not visible in the plot.
}
\label{fig:EP}
\end{figure}

\begin{figure}[htb]
\center
 \includegraphics[width=0.85\linewidth]{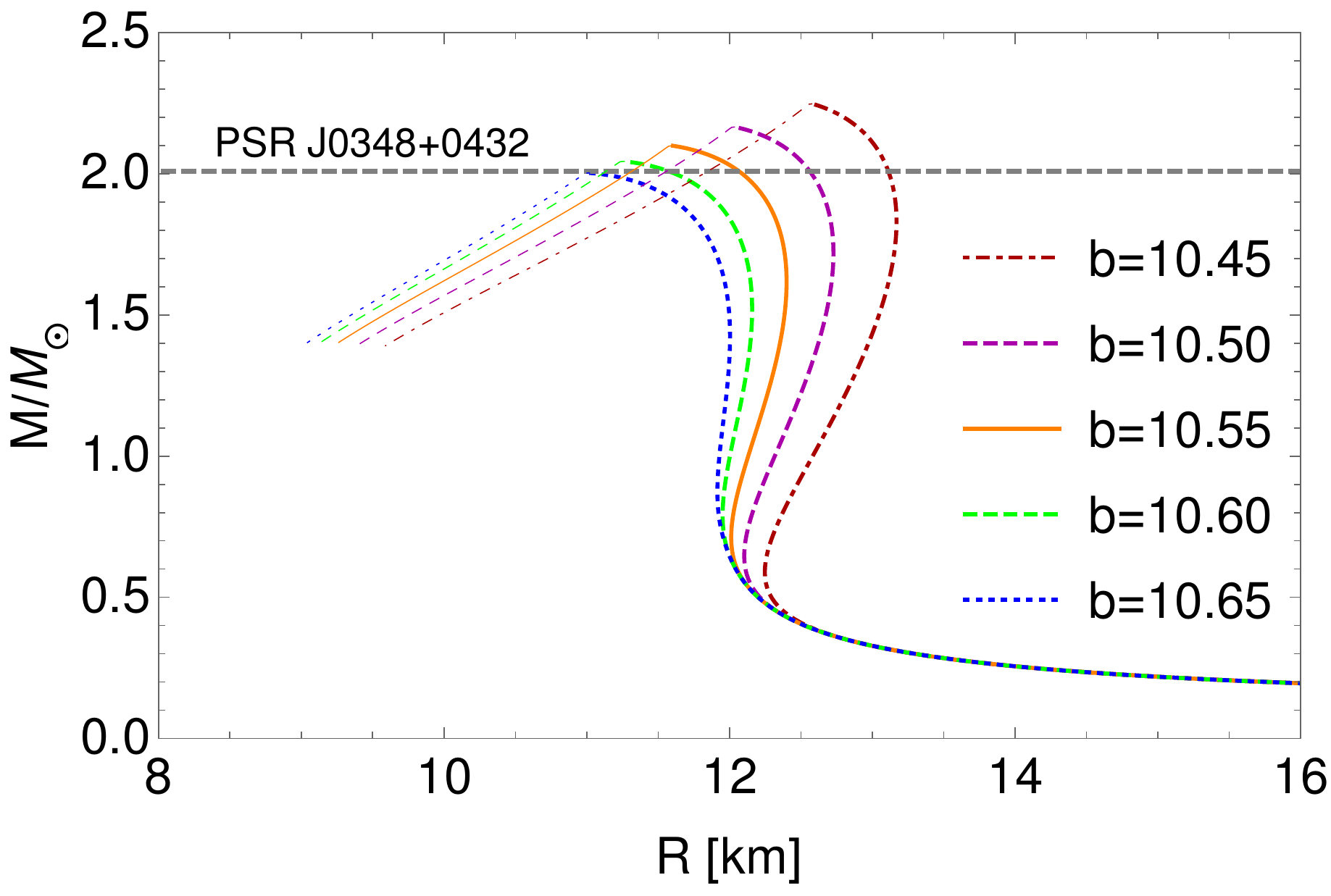}
\caption{Mass-Radius relation for different values of $b$ together with the observational lower mass bound of \cite{Antoniadis:2013pzd}.
} 
\label{fig:MR}
\end{figure}

Basic properties of the obtained hybrid EoSs and transitions  for several choices of $\rho_m$ are listed in Table~\ref{Tab:EOSs1}.
The speed of sound $c_s$ of the holographic EoS depends only weakly on $b$ and is higher than the one of SLy.
Note also that $\rho_m$ depends monotonically on $b$, leading to a more holographic EoS as $b$ decreases.
One can then see in Fig.~\ref{fig:EP} that the higher holographic $c_s$ leads to a higher pressure, and hence increasingly stiff EoSs as $b$ is decreased.
Fig.~\ref{fig:MR} then shows the corresponding mass-radius relation obtained from solving the Tolman-Oppenheimer-Volkoff equations.

Our hybrid EoSs do not support quark matter cores, since the branch with central density larger than $\rho_{\rm QM}$ (thin lines in Fig.~\ref{fig:MR} left of the kinks at the maximum masses) are unstable~\cite{Glendenning:1997wn}. %
Ref.~\cite{Annala:2019puf} argued that NSs without quark matter cores and masses consistent with the current observational bounds are only possible for speeds of sound well above the conformal value, $c_s^2\!>\!1/3$. Our setup provides an explicit realization of such a scenario, as $c_s^2$ is significantly larger than $1/3$ over a wide range of chemical potentials in the holographic baryon phase \cite{Ishii:2019gta}.
Moreover our NS radii are approximately 12--13\,km, which is consistent with recent measurements of X-ray bursts from binary systems~\cite{Nattila:2015jra,Nattila:2017wtj}.

Our choices in Table~\ref{Tab:EOSs1} %
scan over the physically allowed EoSs. Indeed,
the first tabulated hybrid EoS ($b=10.45$) is already too stiff: the value of the tidal deformability $\Lambda \simeq 680$ for a neutron star with mass $\simeq 1.4 M_\odot$ is already well above the bound set by LIGO~\cite{Abbott:2018exr} that $\Lambda \lesssim 580$ (at 90\% confidence level). %
Similarly the maximum mass $M_\mathrm{max}$ for the last hybrid EoS ($b=10.65$) is below the best estimate of the observed mass of the pulsar J0348+0432, $M = 2.01 \pm 0.04\ M_\odot$~\cite{Antoniadis:2013pzd}. %
We choose to use the EoSs with  $b=10.5$ and $b=10.6$, which are consistent with the astrophysical bounds by a clear margin. %

We account for shock-heating effects \cite{PhysRevD.82.084043} during the merger by adding to the cold holographic EoS $p_{\mathrm{c}}(\rho)$ a thermal component $p_{\mathrm{th}}=\Gamma_{\mathrm{th}}\rho
(\epsilon-\epsilon_{\mathrm{c}})$ with $\Gamma_{\mathrm{th}}=1.75$, with $\rho\equiv m_b n_b$ the rest mass density, $m_b$ the baryon mass and $n_b$ the number density.
This gives total pressure $p=p_{\mathrm{c}}+p_{\mathrm{th}}$ and specific internal energy $\epsilon=\epsilon_{\mathrm{c}}+\epsilon_{\mathrm{th}}$.

\noindent
\textbf{4. Numerical Setup.} We use the \texttt{LORENE} pseudo-spectral code \cite{Gourgoulhon:2000nn} to generate initial data for two irrotational stars of equal mass $M$ on quasi-circular orbits with a diameter of $45\,$km, where $M$ is the gravitational mass of the isolated star. These initial data give approximately three ($M=1.5M_\odot$) to six  ($M=1.3M_\odot$) orbits before merger.
The initial data are evolved by solving the $3+1 D$ Einstein equations coupled to ideal general relativistic hydrodynamics (GRHD) \cite{2013rehy.book.....R} using the \texttt{Einstein Toolkit} \cite{Schnetter:2003rb,Thornburg:2003sf,Goodale:2002a}.
\begin{figure*}[ht!]
 \includegraphics[width=0.98\textwidth]{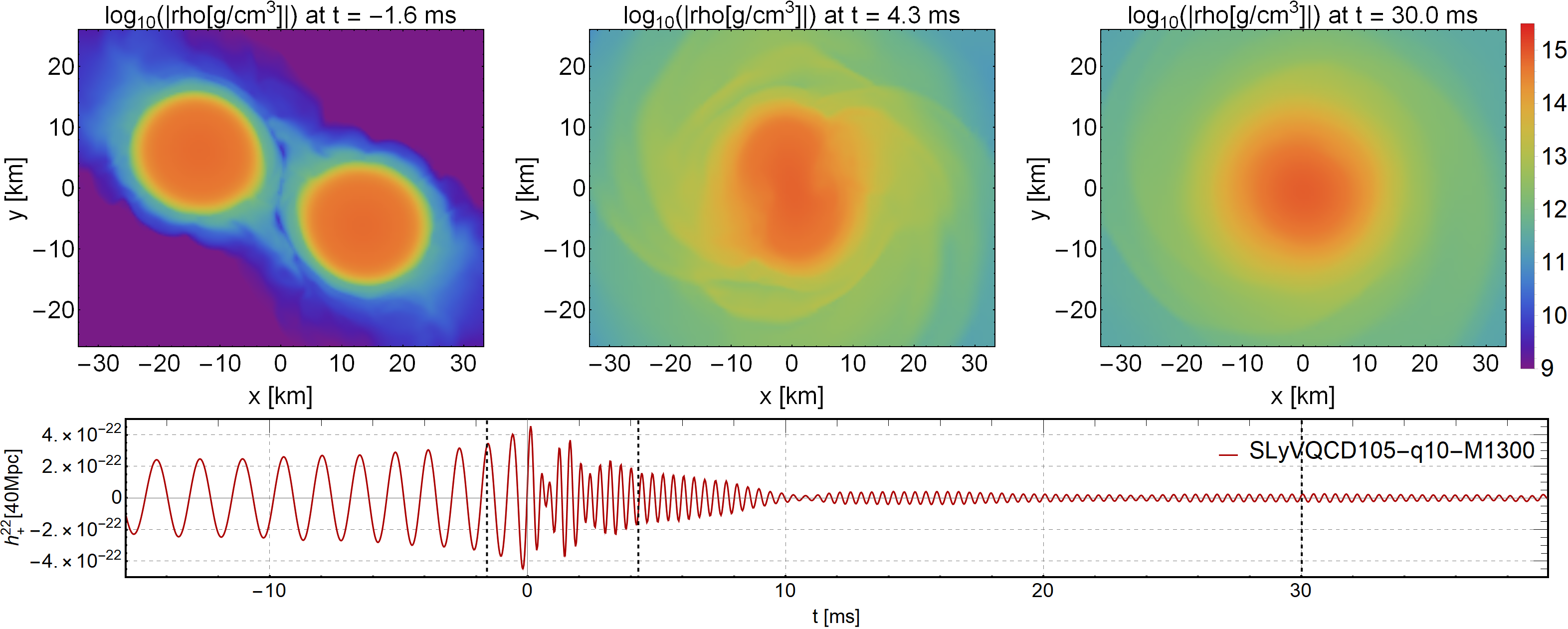}
\caption{Top: Snapshots of the rest mass density in the orbital plane for the late inspiral phase (left), the early post-merger phase (center) and the late post-merger phase (right) for the $M=1.3 M_\odot$ equal mass binary and holographic EoS with $b=10.5$. Bottom: GW strain signal extrapolated to an assumed detector distance of 40Mpc. 
}
\label{fig:densityGW_M1p3}
\end{figure*}

We solve the Einstein equations in \texttt{CCZ4} formulation \cite{PhysRevD.85.064040} using the fourth-order finite differencing code \texttt{McLachlan} \cite{PhysRevD.79.044023,L_ffler_2012} with a ``1+log'' slicing condition and a ``Gamma-driver'' shift condition \cite{PhysRevD.67.084023,PhysRevD.76.124002}.
The GRHD equations are solved in conservative form \cite{Banyuls_1997} using the high-order, high-resolution shock-capturing
code \texttt{WhiskyTHC} \cite{Radice:2012cu,Radice:2013hxh,Radice:2013xpa,Radice:2015nva,refId0}.
We use the method of lines
together with an explicit third-order Runge-Kutta method \cite{Press:2007:NRE:1403886} and prescribe a Courant-Friedrichs-Lewy factor of $0.15$ to compute the step size for the time integration.
All simulations have a volume of size
$\approx3025^3$km$^3$ and assume radiative and static boundary conditions for the metric and the hydrodynamic variables respectively, as well as reflection symmetry across the $z=0$ plane.

Our mesh consists of a central grid (with three refinement levels and resolutions $\Delta h_0\approx 23.5\,$km till $\Delta h_3\approx 2.9\,$km)  and two smaller grids (six refinement levels with finest resolution $\Delta h_6\approx 368\,$m) following the individual stars during inspiral and merger.

\noindent
\textbf{5. Waveform Extraction.} 
To compute the GW signal we use the Newman-Penrose formalism \cite{article} in which the GW polarization amplitudes $h_{+,\times}$ are related to the Weyl curvature scalar $\psi_4$ via 
\begin{equation}
\ddot{h}_+ -i\ddot{h}_\times=\psi_4=\sum_{\ell=2}^\infty \sum_{m=-\ell}^{\ell}\psi_4^{\ell m} {_{-2}Y_{\ell m}}(\theta,\varphi)\,,
\end{equation}
where dot means time derivative and $_{s}Y_{\ell m}(\theta,\varphi)$ are spin-weighted spherical harmonics of weight $s\!=\!-2$.
In the simulation we extract $\psi_4$ at a spherical surface with radius  $R\!\approx\! 884\,$km at a sampling rate of $\approx\! 169$\,kHz.
In the following we will only consider the dominant $\ell\!=\!m\!=\!2$ modes $h_{+,\times}^{2 2}$
and assume optimal orientation ($_{-2}Y_{2 2}(0,0)=\frac{1}{2}\sqrt{5/\pi}$) of the merger with respect to the detector.
We then extrapolate the signal to 40\,Mpc, i.e. the estimated luminosity distance of GW170817 \cite{TheLIGOScientific:2017qsa}.
To analyze the spectral features of the waveform we follow \cite{Takami:2014zpa} and compute the PSD defined as
\begin{equation}\label{eq:PSD}
\tilde{h}(f)\!\equiv\!\sqrt{\frac{|\!\int\! h_{+}(t)e^{-i2\pi f t}dt|^2\!+\!|\!\int\! h_{\times}(t)e^{-i2\pi f t}dt|^2}{2}},
\end{equation}
where we take the time interval between $-7$ and $24\,$ms and define $t=0$ by the maximum of the GW amplitude.

\begin{figure*}[ht!]
 \includegraphics[width=0.98\textwidth]{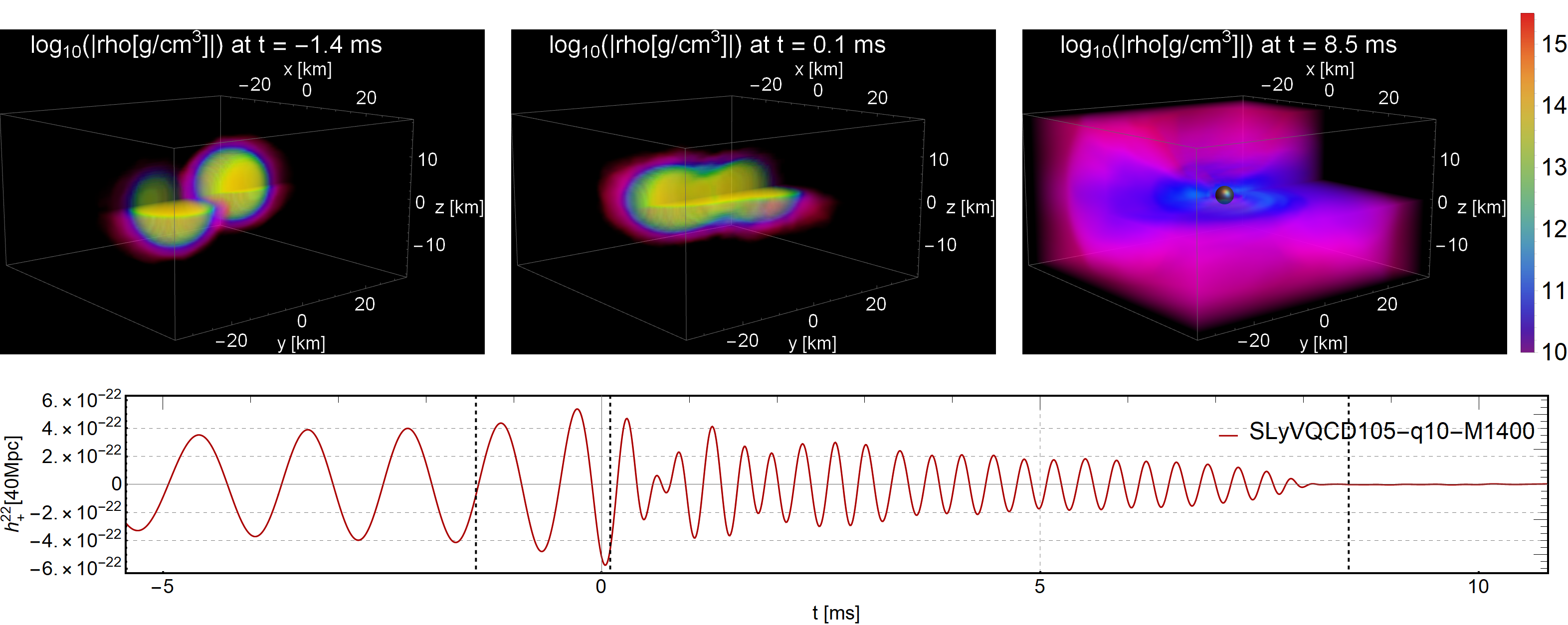}
\caption{As in Fig.~\ref{fig:densityGW_M1p3}, but shown here for $M=1.4 M_\odot$.
Snapshots are at the inspiral (left), merger (center) and after black hole collapse (right). For visibility we omit the densities in one quadrant.
}
\label{fig:densityGW_M1p4}
\end{figure*}

\begin{figure}[h]
 \includegraphics[width=0.4\textwidth]{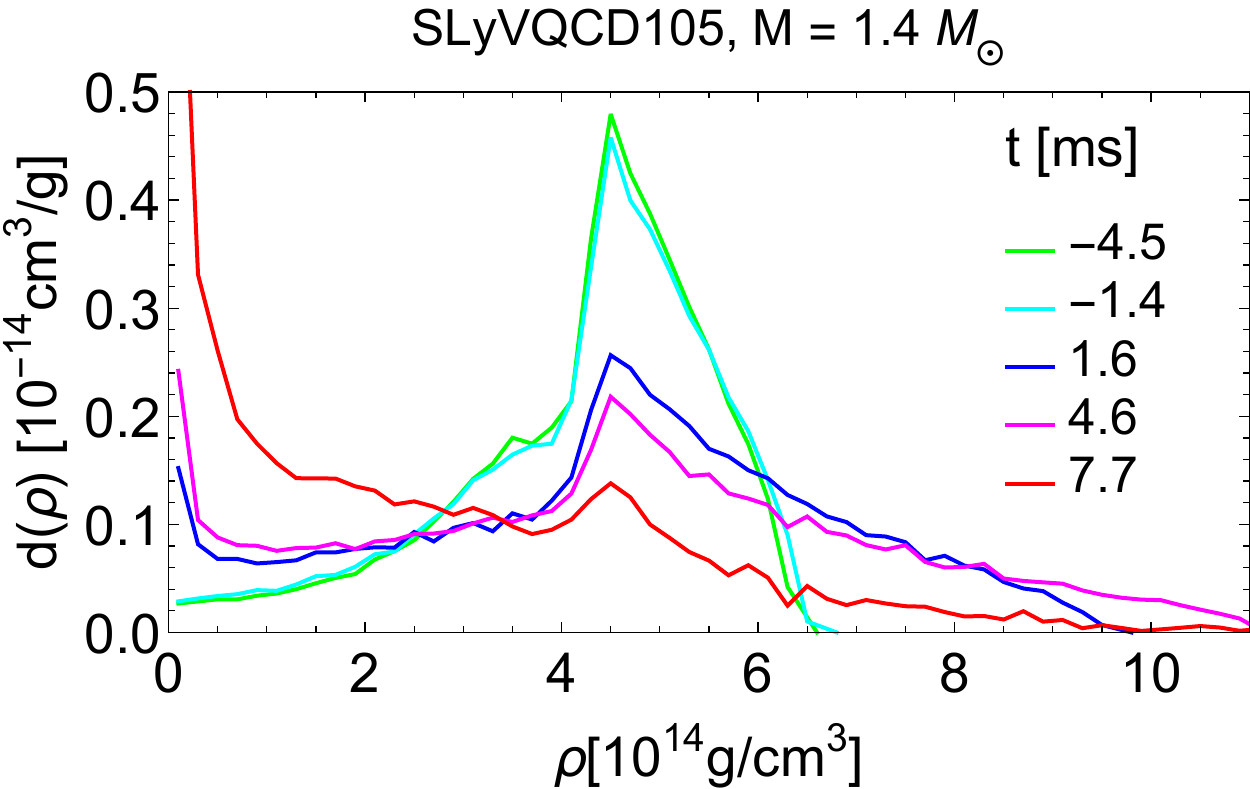}
\caption{Distribution of the $\rho$ field of Fig. \ref{fig:densityGW_M1p4} at different times. After the merger the HMNS reaches a density of about twice the maximum density of the initial star. Shortly after the black hole forms ($t\!\approx\!7.8$ms) most matter is in the black hole with some low density matter in the surrounding atmosphere.}
\label{fig:densityPDF}
\end{figure}

\noindent
\textbf{6. Results.}
In Fig.~\ref{fig:densityGW_M1p3} we show snapshots of the rest mass density $\rho$ in the orbital plane and the corresponding waveform for a binary merger with individual mass $M\!=\!1.3M_\odot$ for our hybrid EoS with $b\!=\!10.5$ 
\footnote{Movies of the mergers can be found at 
\href{https://sites.google.com/site/wilkevanderschee/neutron-stars}{sites.google.com/site/wilkevanderschee/neutron-stars}}.
The first plot ($t\!=\!-1.6$\,ms)  
shows a late stage of the inspiral, where the mutual tidal deformations of the stars become significant and full numerical GR simulations become important. The second plot ($t\!=\!4.3$\,ms) shows an early post-merger stage, where the rotating and oscillating HMNS is bar-deformed and still non-axially symmetric.
This phase lasts for about $10$ms and the corresponding GW signal contains spectral features that are sensitive to the EoS, which we shall analyze below.
The third plot ($t\!=\!30$\,ms) shows the late post-merger stage, where the merger remnant has settled down to an almost axially symmetric rotating HMNS with a lifetime $>\!40$\,ms.
The $M\!=\!1.4M_\odot$ case (Fig.~\ref{fig:densityGW_M1p4}) results in a relatively short-lived, highly deformed HMNS, which collapses $\approx\!7.8$\,ms after the merger into a black hole with a torus of remaining in-falling matter.
Fig. \ref{fig:densityPDF} shows the distribution of densities $d(\rho,\,t)$ present in the star for several times, defined as $\int d(\rho,\,t) \dd \rho = 1$, where $d(\rho,\,t)\dd\rho$ is the fraction of the star that has a density between $\rho$ and $\rho+\dd\rho$.
After the merger the HMNS reaches almost twice the maximum rest-mass density of the original stars.
This density is still not enough to reach the quark matter phase (see Fig.~\ref{fig:EP}), so unfortunately neutron star mergers do not explore the phase transition of our hybrid EoSs. The heavier cases ($M\geq1.5 M_\odot$) lead to immediate collapse after the merger and give no HMNS phase.

\begin{figure*}[ht!]
 \includegraphics[width=0.47\textwidth]{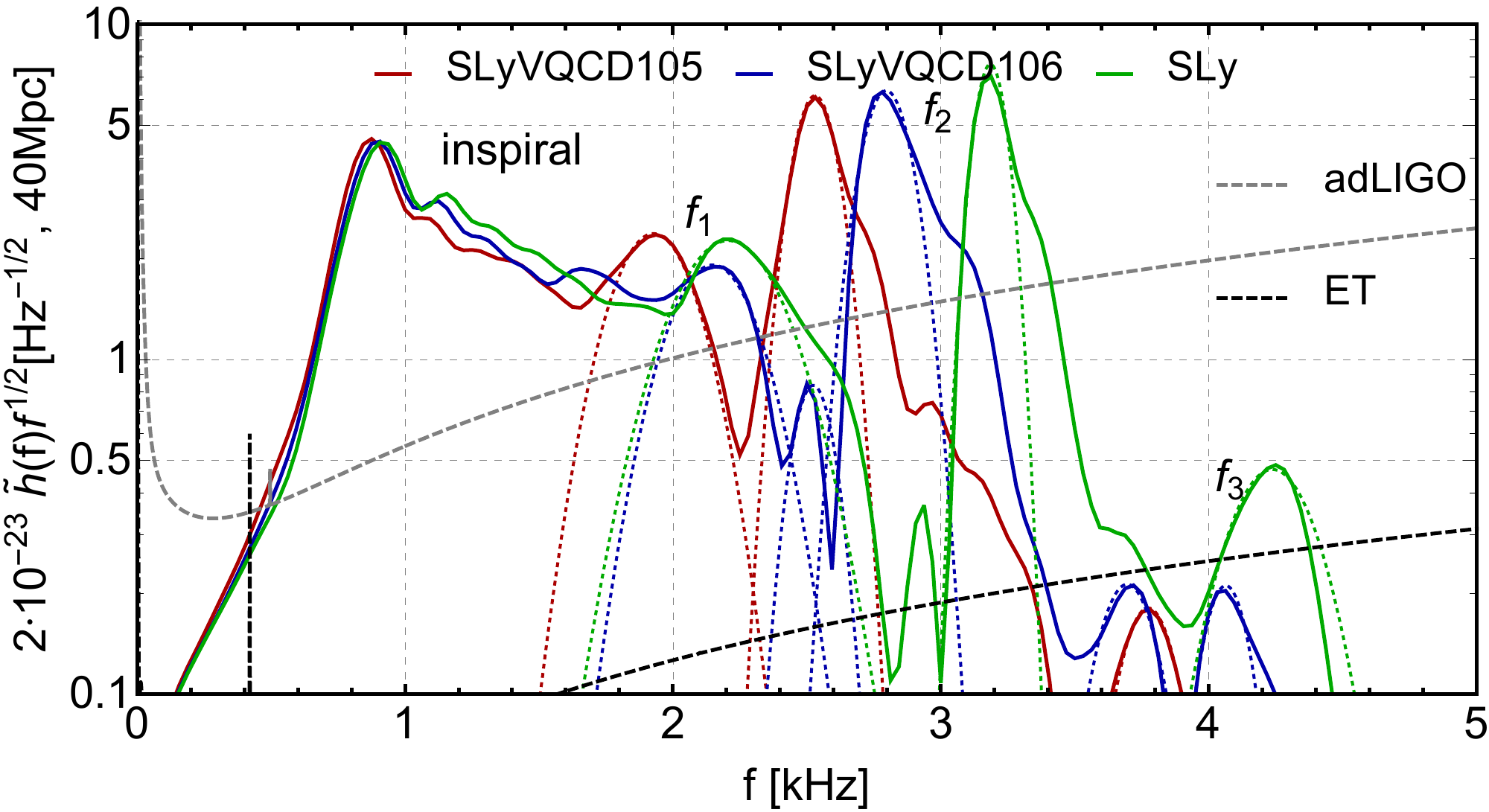}
 \includegraphics[width=0.47\textwidth]{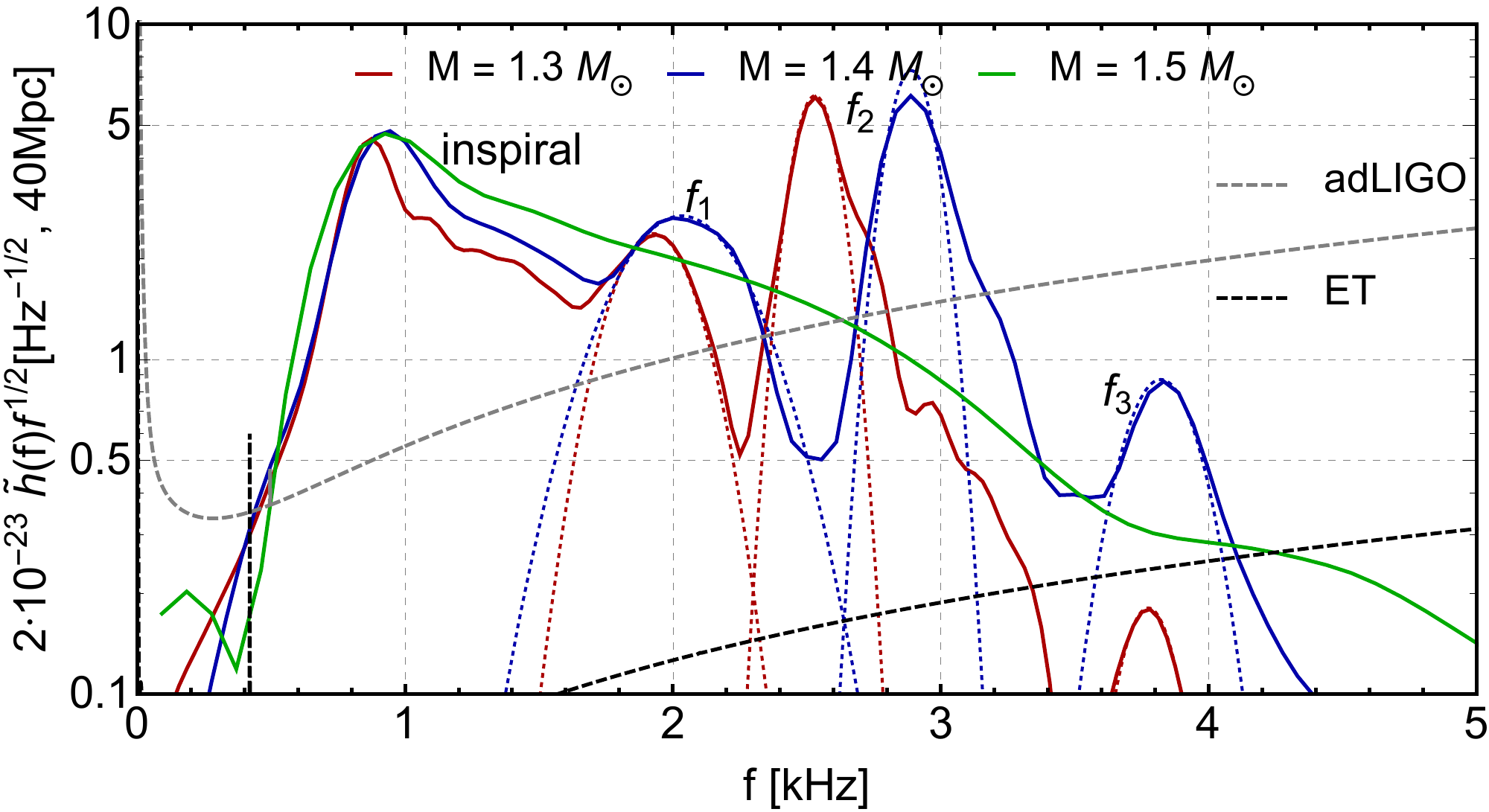}
\caption{
We show the power spectrum density (PSD) for two holographic EOSs with a comparison to the SLy EOS for mergers with $M=1.3 M_\odot$ (left) and a comparison of different masses for our holographic EOS with $b\!=\!10.5$ (right). The Gaussian fits (dotted) determine the characteristic frequencies $f_1$, $f_2$ and $f_3$ in the post-merger phase, except when $M = 1.5 M_\odot$, which collapses to a black hole immediately.
Black and gray dashed lines are sensitivity curves of the advanced LIGO (adLIGO) detector \cite{adLIGO2018} and the Einstein Telescope (ET) \cite{Punturo:2010zza}.}
\label{fig:PSD}
\end{figure*}

In Fig.~\ref{fig:PSD} we show the PSDs (Eqn.~(\ref{eq:PSD})) of our numerical waveforms together with the sensitivity curves of Advanced LIGO (adLIGO) \cite{adLIGO2018} and of the Einstein Telescope (ET) \cite{Punturo:2010zza}.
These spectra show several pronounced features, where the lowest frequency peak can be attributed to the inspiral phase and the three higher frequency peaks ($f_1$,$f_2$,$f_3$) to the post-merger phase (for details see \cite{Baiotti:2016qnr}).
In the left plot we keep $M=1.3M_\odot$ fixed and show our results from two different holographic EoSs with $b\!=\!10.5$ (red) and $b\!=\!10.6$ (blue) together with the result for standard SLy (green) without input from holography.
Increasing $b$ shifts the post-merger peaks towards the higher SLy frequencies, since higher $b$ has a higher matching density (see Table \ref{Tab:EOSs1}).
This behavior is in agreement with the universality of $f_1$ as function of the compactness ($M/R$) found in \cite{Takami:2014zpa}, which predicts a shift to higher frequencies for softer EoSs 
(increasing $b$ in our case) giving more compact stars (see Fig.~\ref{fig:MR}).
The picture is less clear for the highest frequency peak $f_3$, which in some cases is ambiguous. In contrast to the SLy EoS, we find that the $f_3$ peak predicted from our holographic EoS will probably not be visible to third generation detectors for $M\!=\!1.3\,M_\odot$ binaries.
In Fig.~\ref{fig:PSD} (right) we study the PSD for fixed holographic EoS ($b\!=\!10.5$) and three different binary masses. The $M\!=\!1.5\,M_\odot$ case leads to immediate collapse without HMNS phase and thus to a featureless PSD. 
For increasing mass, which in the considered mass range results in an increase of compactness, the characteristic frequencies are shifted to higher values.
Furthermore the $M=1.4\,M_\odot$ case gives, in contrast to $M=1.3\,M_\odot$, a distinguished $f_3$ peak that lies clearly above the sensitivity curve of the ET.

Finally in Table \ref{Tab:simulations} we list the values for $f_1,f_2$ and $f_3$ we extracted from our simulations using a Gaussian fit of the eight frequency points around the local maxima of the peaks.
The values of $f_1$ we find for the pure SLy and our hybrid EoS with $b=10.5$ are in good agreement with the universal curve proposed in \cite{Takami:2014zpa}. 
\begin{table}[htb]
 \caption{Summary of simulation results. Ambiguous frequencies are written in brackets.}
 \begin{tabular}{ccccccc}\hline\hline
  $M$[$M_\odot$] & EoS & $b$ & $f_1$[kHz] & $f_2$[kHz] & $f_3$[kHz] \\
  \hline\hline
 1.30 & SLyVQCD105  & 10.5 &  1.93  & 2.53  & 3.77 \\
 1.30 & SLyVQCD106  & 10.6 &  2.15  & 2.80 & 3.70 (4.06) \\
 1.30 & SLy         &  -   &  2.21 & 3.19 & 4.24 \\
 1.35 & SLyVQCD105  & 10.5 &  1.95 & 2.60  & 3.53 (3.90)\\
 1.40 & SLyVQCD105  & 10.5 &  2.03 & 2.89  & 3.82 \\
 1.50 & SLyVQCD105  & 10.5 &  -- & -- & -- \\
 \hline
\end{tabular}
\label{Tab:simulations} 
\end{table}

\noindent
\textbf{7. Discussion.}
The characteristic frequencies ($f_1,f_2,f_3$) of our holographic EoS from  the HMNS phase of the neutron star mergers establish a novel approach to confront predictions from holography to future observational data.
Our results from the holographic model predict a shift of $f_1$ and $f_2$ to significantly lower frequencies compared to the non-holographic SLy model, which can be attributed to the higher stiffness of our hybrid EoS. Furthermore the holographic EoS predicts that $f_3$ is best visible for masses  around $1.4\,M_\odot$, as its amplitude is too low for lower masses, and higher masses lead to quick black hole formation. In our simulations it is not possible to reach the phase transition to the quark phase, 
as the highest attained density for binaries that do not immediately  collapse is about $1.1 \cdot 10^{15}$g/cm$^3$ (Fig. \ref{fig:densityPDF}).

The present work can be seen as a first step towards holographic GW model building and is extendable in many different ways.
Important next steps are to improve the holographic model by going beyond the homogeneous approach for the V-QCD baryon field and also to systematically study other nuclear matter models than SLy used for the low density part in this work.
Due to limited computing resources we use rather coarse meshes in our simulations, which is something we plan to improve on in future works.
More challenging extensions are to include finite temperature effects in the EoS and magnetic fields in the merger simulations. It would also be exciting to study neutrinos and electromagnetic radiation 
in this current age of multi-messenger astronomy.

\noindent
\textbf{Acknowledgements - }
It is a pleasure to thank Umut G\"ursoy, Takaaki Ishii, Aleksi Kurkela, Raimond Snellings and Stefan Vandoren for interesting discussions. CE thanks Niko Jokela and Aleksi Vuorinen for motivating this project.  We especially thank Elias Most for introducing us to \texttt{WhiskyTHC} and Helvi Witek for invaluable support with the \texttt{Einstein Toolkit}.
WS gratefully acknowledges the hospitality of the CERN Theory group.
This work is 
partially supported by the Netherlands Organisation for Scientific Research (NWO)
under the grants 680-47-518 (Vidi), 680-47-458 (WS, Veni) and the Delta-Institute for Theoretical Physics ($\Delta$-ITP), both funded by the Dutch Ministry of Education, Culture and Science (OCW).
 This work was carried out on the Dutch national e-infrastructure with the support of SURF Cooperative. 

\bibliographystyle{apsrev4-1}
\bibliography{biblio}

\begin{thebibliography}{59}%
\makeatletter
\providecommand \@ifxundefined [1]{%
 \@ifx{#1\undefined}
}%
\providecommand \@ifnum [1]{%
 \ifnum #1\expandafter \@firstoftwo
 \else \expandafter \@secondoftwo
 \fi
}%
\providecommand \@ifx [1]{%
 \ifx #1\expandafter \@firstoftwo
 \else \expandafter \@secondoftwo
 \fi
}%
\providecommand \natexlab [1]{#1}%
\providecommand \enquote  [1]{``#1''}%
\providecommand \bibnamefont  [1]{#1}%
\providecommand \bibfnamefont [1]{#1}%
\providecommand \citenamefont [1]{#1}%
\providecommand \href@noop [0]{\@secondoftwo}%
\providecommand \href [0]{\begingroup \@sanitize@url \@href}%
\providecommand \@href[1]{\@@startlink{#1}\@@href}%
\providecommand \@@href[1]{\endgroup#1\@@endlink}%
\providecommand \@sanitize@url [0]{\catcode `\\12\catcode `\$12\catcode
  `\&12\catcode `\#12\catcode `\^12\catcode `\_12\catcode `\%12\relax}%
\providecommand \@@startlink[1]{}%
\providecommand \@@endlink[0]{}%
\providecommand \url  [0]{\begingroup\@sanitize@url \@url }%
\providecommand \@url [1]{\endgroup\@href {#1}{\urlprefix }}%
\providecommand \urlprefix  [0]{URL }%
\providecommand \Eprint [0]{\href }%
\providecommand \doibase [0]{http://dx.doi.org/}%
\providecommand \selectlanguage [0]{\@gobble}%
\providecommand \bibinfo  [0]{\@secondoftwo}%
\providecommand \bibfield  [0]{\@secondoftwo}%
\providecommand \translation [1]{[#1]}%
\providecommand \BibitemOpen [0]{}%
\providecommand \bibitemStop [0]{}%
\providecommand \bibitemNoStop [0]{.\EOS\space}%
\providecommand \EOS [0]{\spacefactor3000\relax}%
\providecommand \BibitemShut  [1]{\csname bibitem#1\endcsname}%
\let\auto@bib@innerbib\@empty
\bibitem [{\citenamefont {Abbott}\ \emph {et~al.}(2016)\citenamefont {Abbott}
  \emph {et~al.}}]{Abbott:2016blz}%
  \BibitemOpen
  \bibfield  {author} {\bibinfo {author} {\bibfnamefont {B.~P.}\ \bibnamefont
  {Abbott}} \emph {et~al.} (\bibinfo {collaboration} {LIGO Scientific,
  Virgo}),\ }\href {\doibase 10.1103/PhysRevLett.116.061102} {\bibfield
  {journal} {\bibinfo  {journal} {Phys. Rev. Lett.}\ }\textbf {\bibinfo
  {volume} {116}},\ \bibinfo {pages} {061102} (\bibinfo {year} {2016})},\
  \Eprint {http://arxiv.org/abs/1602.03837} {arXiv:1602.03837 [gr-qc]}
  \BibitemShut {NoStop}%
\bibitem [{\citenamefont {Abbott}\ \emph {et~al.}(2017)\citenamefont {Abbott}
  \emph {et~al.}}]{TheLIGOScientific:2017qsa}%
  \BibitemOpen
  \bibfield  {author} {\bibinfo {author} {\bibfnamefont {B.}~\bibnamefont
  {Abbott}} \emph {et~al.} (\bibinfo {collaboration} {LIGO Scientific,
  Virgo}),\ }\href {\doibase 10.1103/PhysRevLett.119.161101} {\bibfield
  {journal} {\bibinfo  {journal} {Phys. Rev. Lett.}\ }\textbf {\bibinfo
  {volume} {119}},\ \bibinfo {pages} {161101} (\bibinfo {year} {2017})},\
  \Eprint {http://arxiv.org/abs/1710.05832} {arXiv:1710.05832 [gr-qc]}
  \BibitemShut {NoStop}%
\bibitem [{\citenamefont {Baiotti}\ and\ \citenamefont
  {Rezzolla}(2017)}]{Baiotti:2016qnr}%
  \BibitemOpen
  \bibfield  {author} {\bibinfo {author} {\bibfnamefont {L.}~\bibnamefont
  {Baiotti}}\ and\ \bibinfo {author} {\bibfnamefont {L.}~\bibnamefont
  {Rezzolla}},\ }\href {\doibase 10.1088/1361-6633/aa67bb} {\bibfield
  {journal} {\bibinfo  {journal} {Rept. Prog. Phys.}\ }\textbf {\bibinfo
  {volume} {80}},\ \bibinfo {pages} {096901} (\bibinfo {year} {2017})},\
  \Eprint {http://arxiv.org/abs/1607.03540} {arXiv:1607.03540 [gr-qc]}
  \BibitemShut {NoStop}%
\bibitem [{\citenamefont {Gandolfi}\ \emph {et~al.}(2012)\citenamefont
  {Gandolfi}, \citenamefont {Carlson},\ and\ \citenamefont
  {Reddy}}]{Gandolfi:2011xu}%
  \BibitemOpen
  \bibfield  {author} {\bibinfo {author} {\bibfnamefont {S.}~\bibnamefont
  {Gandolfi}}, \bibinfo {author} {\bibfnamefont {J.}~\bibnamefont {Carlson}}, \
  and\ \bibinfo {author} {\bibfnamefont {S.}~\bibnamefont {Reddy}},\ }\href
  {\doibase 10.1103/PhysRevC.85.032801} {\bibfield  {journal} {\bibinfo
  {journal} {Phys. Rev.}\ }\textbf {\bibinfo {volume} {C85}},\ \bibinfo {pages}
  {032801} (\bibinfo {year} {2012})},\ \Eprint {http://arxiv.org/abs/1101.1921}
  {arXiv:1101.1921 [nucl-th]} \BibitemShut {NoStop}%
\bibitem [{\citenamefont {Gorda}\ \emph {et~al.}(2018)\citenamefont {Gorda},
  \citenamefont {Kurkela}, \citenamefont {Romatschke}, \citenamefont {Säppi},\
  and\ \citenamefont {Vuorinen}}]{Gorda:2018gpy}%
  \BibitemOpen
  \bibfield  {author} {\bibinfo {author} {\bibfnamefont {T.}~\bibnamefont
  {Gorda}}, \bibinfo {author} {\bibfnamefont {A.}~\bibnamefont {Kurkela}},
  \bibinfo {author} {\bibfnamefont {P.}~\bibnamefont {Romatschke}}, \bibinfo
  {author} {\bibfnamefont {M.}~\bibnamefont {Säppi}}, \ and\ \bibinfo {author}
  {\bibfnamefont {A.}~\bibnamefont {Vuorinen}},\ }\href {\doibase
  10.1103/PhysRevLett.121.202701} {\bibfield  {journal} {\bibinfo  {journal}
  {Phys. Rev. Lett.}\ }\textbf {\bibinfo {volume} {121}},\ \bibinfo {pages}
  {202701} (\bibinfo {year} {2018})},\ \Eprint
  {http://arxiv.org/abs/1807.04120} {arXiv:1807.04120 [hep-ph]} \BibitemShut
  {NoStop}%
\bibitem [{\citenamefont {Annala}\ \emph {et~al.}(2019)\citenamefont {Annala},
  \citenamefont {Gorda}, \citenamefont {Kurkela}, \citenamefont {Nättilä},\
  and\ \citenamefont {Vuorinen}}]{Annala:2019puf}%
  \BibitemOpen
  \bibfield  {author} {\bibinfo {author} {\bibfnamefont {E.}~\bibnamefont
  {Annala}}, \bibinfo {author} {\bibfnamefont {T.}~\bibnamefont {Gorda}},
  \bibinfo {author} {\bibfnamefont {A.}~\bibnamefont {Kurkela}}, \bibinfo
  {author} {\bibfnamefont {J.}~\bibnamefont {Nättilä}}, \ and\ \bibinfo
  {author} {\bibfnamefont {A.}~\bibnamefont {Vuorinen}},\ }\href@noop {} {\
  (\bibinfo {year} {2019})},\ \Eprint {http://arxiv.org/abs/1903.09121}
  {arXiv:1903.09121 [astro-ph.HE]} \BibitemShut {NoStop}%
\bibitem [{\citenamefont {Annala}\ \emph {et~al.}(2018)\citenamefont {Annala},
  \citenamefont {Gorda}, \citenamefont {Kurkela},\ and\ \citenamefont
  {Vuorinen}}]{Annala:2017llu}%
  \BibitemOpen
  \bibfield  {author} {\bibinfo {author} {\bibfnamefont {E.}~\bibnamefont
  {Annala}}, \bibinfo {author} {\bibfnamefont {T.}~\bibnamefont {Gorda}},
  \bibinfo {author} {\bibfnamefont {A.}~\bibnamefont {Kurkela}}, \ and\
  \bibinfo {author} {\bibfnamefont {A.}~\bibnamefont {Vuorinen}},\ }\href
  {\doibase 10.1103/PhysRevLett.120.172703} {\bibfield  {journal} {\bibinfo
  {journal} {Phys. Rev. Lett.}\ }\textbf {\bibinfo {volume} {120}},\ \bibinfo
  {pages} {172703} (\bibinfo {year} {2018})},\ \Eprint
  {http://arxiv.org/abs/1711.02644} {arXiv:1711.02644 [astro-ph.HE]}
  \BibitemShut {NoStop}%
\bibitem [{\citenamefont {Most}\ \emph
  {et~al.}(2018{\natexlab{a}})\citenamefont {Most}, \citenamefont {Weih},
  \citenamefont {Rezzolla},\ and\ \citenamefont
  {Schaffner-Bielich}}]{Most:2018hfd}%
  \BibitemOpen
  \bibfield  {author} {\bibinfo {author} {\bibfnamefont {E.~R.}\ \bibnamefont
  {Most}}, \bibinfo {author} {\bibfnamefont {L.~R.}\ \bibnamefont {Weih}},
  \bibinfo {author} {\bibfnamefont {L.}~\bibnamefont {Rezzolla}}, \ and\
  \bibinfo {author} {\bibfnamefont {J.}~\bibnamefont {Schaffner-Bielich}},\
  }\href {\doibase 10.1103/PhysRevLett.120.261103} {\bibfield  {journal}
  {\bibinfo  {journal} {Phys. Rev. Lett.}\ }\textbf {\bibinfo {volume} {120}},\
  \bibinfo {pages} {261103} (\bibinfo {year} {2018}{\natexlab{a}})},\ \Eprint
  {http://arxiv.org/abs/1803.00549} {arXiv:1803.00549 [gr-qc]} \BibitemShut
  {NoStop}%
\bibitem [{\citenamefont {Most}\ \emph
  {et~al.}(2018{\natexlab{b}})\citenamefont {Most}, \citenamefont {Papenfort},
  \citenamefont {Dexheimer}, \citenamefont {Hanauske}, \citenamefont {Schramm},
  \citenamefont {Stöcker},\ and\ \citenamefont {Rezzolla}}]{Most:2018eaw}%
  \BibitemOpen
  \bibfield  {author} {\bibinfo {author} {\bibfnamefont {E.~R.}\ \bibnamefont
  {Most}}, \bibinfo {author} {\bibfnamefont {L.~J.}\ \bibnamefont {Papenfort}},
  \bibinfo {author} {\bibfnamefont {V.}~\bibnamefont {Dexheimer}}, \bibinfo
  {author} {\bibfnamefont {M.}~\bibnamefont {Hanauske}}, \bibinfo {author}
  {\bibfnamefont {S.}~\bibnamefont {Schramm}}, \bibinfo {author} {\bibfnamefont
  {H.}~\bibnamefont {Stöcker}}, \ and\ \bibinfo {author} {\bibfnamefont
  {L.}~\bibnamefont {Rezzolla}},\ }\href@noop {} {\  (\bibinfo {year}
  {2018}{\natexlab{b}})},\ \Eprint {http://arxiv.org/abs/1807.03684}
  {arXiv:1807.03684 [astro-ph.HE]} \BibitemShut {NoStop}%
\bibitem [{\citenamefont {Bauswein}\ \emph {et~al.}(2018)\citenamefont
  {Bauswein}, \citenamefont {Bastian}, \citenamefont {Blaschke}, \citenamefont
  {Chatziioannou}, \citenamefont {Clark}, \citenamefont {Fischer},\ and\
  \citenamefont {Oertel}}]{Bauswein:2018bma}%
  \BibitemOpen
  \bibfield  {author} {\bibinfo {author} {\bibfnamefont {A.}~\bibnamefont
  {Bauswein}}, \bibinfo {author} {\bibfnamefont {N.-U.~F.}\ \bibnamefont
  {Bastian}}, \bibinfo {author} {\bibfnamefont {D.~B.}\ \bibnamefont
  {Blaschke}}, \bibinfo {author} {\bibfnamefont {K.}~\bibnamefont
  {Chatziioannou}}, \bibinfo {author} {\bibfnamefont {J.~A.}\ \bibnamefont
  {Clark}}, \bibinfo {author} {\bibfnamefont {T.}~\bibnamefont {Fischer}}, \
  and\ \bibinfo {author} {\bibfnamefont {M.}~\bibnamefont {Oertel}},\
  }\href@noop {} {\  (\bibinfo {year} {2018})},\ \Eprint
  {http://arxiv.org/abs/1809.01116} {arXiv:1809.01116 [astro-ph.HE]}
  \BibitemShut {NoStop}%
\bibitem [{\citenamefont {Loffler}\ \emph {et~al.}(2012)\citenamefont {Loffler}
  \emph {et~al.}}]{Loffler:2011ay}%
  \BibitemOpen
  \bibfield  {author} {\bibinfo {author} {\bibfnamefont {F.}~\bibnamefont
  {Loffler}} \emph {et~al.},\ }\href {\doibase 10.1088/0264-9381/29/11/115001}
  {\bibfield  {journal} {\bibinfo  {journal} {Class. Quant. Grav.}\ }\textbf
  {\bibinfo {volume} {29}},\ \bibinfo {pages} {115001} (\bibinfo {year}
  {2012})},\ \Eprint {http://arxiv.org/abs/1111.3344} {arXiv:1111.3344 [gr-qc]}
  \BibitemShut {NoStop}%
\bibitem [{\citenamefont {Takami}\ \emph {et~al.}(2014)\citenamefont {Takami},
  \citenamefont {Rezzolla},\ and\ \citenamefont {Baiotti}}]{Takami:2014zpa}%
  \BibitemOpen
  \bibfield  {author} {\bibinfo {author} {\bibfnamefont {K.}~\bibnamefont
  {Takami}}, \bibinfo {author} {\bibfnamefont {L.}~\bibnamefont {Rezzolla}}, \
  and\ \bibinfo {author} {\bibfnamefont {L.}~\bibnamefont {Baiotti}},\ }\href
  {\doibase 10.1103/PhysRevLett.113.091104} {\bibfield  {journal} {\bibinfo
  {journal} {Phys. Rev. Lett.}\ }\textbf {\bibinfo {volume} {113}},\ \bibinfo
  {pages} {091104} (\bibinfo {year} {2014})},\ \Eprint
  {http://arxiv.org/abs/1403.5672} {arXiv:1403.5672 [gr-qc]} \BibitemShut
  {NoStop}%
\bibitem [{\citenamefont {Takami}\ \emph {et~al.}(2015)\citenamefont {Takami},
  \citenamefont {Rezzolla},\ and\ \citenamefont {Baiotti}}]{Takami:2014tva}%
  \BibitemOpen
  \bibfield  {author} {\bibinfo {author} {\bibfnamefont {K.}~\bibnamefont
  {Takami}}, \bibinfo {author} {\bibfnamefont {L.}~\bibnamefont {Rezzolla}}, \
  and\ \bibinfo {author} {\bibfnamefont {L.}~\bibnamefont {Baiotti}},\ }\href
  {\doibase 10.1103/PhysRevD.91.064001} {\bibfield  {journal} {\bibinfo
  {journal} {Phys. Rev.}\ }\textbf {\bibinfo {volume} {D91}},\ \bibinfo {pages}
  {064001} (\bibinfo {year} {2015})},\ \Eprint {http://arxiv.org/abs/1412.3240}
  {arXiv:1412.3240 [gr-qc]} \BibitemShut {NoStop}%
\bibitem [{\citenamefont {Radice}\ \emph {et~al.}(2017)\citenamefont {Radice},
  \citenamefont {Bernuzzi}, \citenamefont {Del~Pozzo}, \citenamefont
  {Roberts},\ and\ \citenamefont {Ott}}]{Radice:2016rys}%
  \BibitemOpen
  \bibfield  {author} {\bibinfo {author} {\bibfnamefont {D.}~\bibnamefont
  {Radice}}, \bibinfo {author} {\bibfnamefont {S.}~\bibnamefont {Bernuzzi}},
  \bibinfo {author} {\bibfnamefont {W.}~\bibnamefont {Del~Pozzo}}, \bibinfo
  {author} {\bibfnamefont {L.~F.}\ \bibnamefont {Roberts}}, \ and\ \bibinfo
  {author} {\bibfnamefont {C.~D.}\ \bibnamefont {Ott}},\ }\href {\doibase
  10.3847/2041-8213/aa775f} {\bibfield  {journal} {\bibinfo  {journal}
  {Astrophys. J.}\ }\textbf {\bibinfo {volume} {842}},\ \bibinfo {pages} {L10}
  (\bibinfo {year} {2017})},\ \Eprint {http://arxiv.org/abs/1612.06429}
  {arXiv:1612.06429 [astro-ph.HE]} \BibitemShut {NoStop}%
\bibitem [{\citenamefont {Maione}\ \emph {et~al.}(2017)\citenamefont {Maione},
  \citenamefont {De~Pietri}, \citenamefont {Feo},\ and\ \citenamefont
  {Löffler}}]{Maione:2017aux}%
  \BibitemOpen
  \bibfield  {author} {\bibinfo {author} {\bibfnamefont {F.}~\bibnamefont
  {Maione}}, \bibinfo {author} {\bibfnamefont {R.}~\bibnamefont {De~Pietri}},
  \bibinfo {author} {\bibfnamefont {A.}~\bibnamefont {Feo}}, \ and\ \bibinfo
  {author} {\bibfnamefont {F.}~\bibnamefont {Löffler}},\ }\href {\doibase
  10.1103/PhysRevD.96.063011} {\bibfield  {journal} {\bibinfo  {journal} {Phys.
  Rev.}\ }\textbf {\bibinfo {volume} {D96}},\ \bibinfo {pages} {063011}
  (\bibinfo {year} {2017})},\ \Eprint {http://arxiv.org/abs/1707.03368}
  {arXiv:1707.03368 [gr-qc]} \BibitemShut {NoStop}%
\bibitem [{\citenamefont {Järvinen}\ and\ \citenamefont
  {Kiritsis}(2012)}]{Jarvinen:2011qe}%
  \BibitemOpen
  \bibfield  {author} {\bibinfo {author} {\bibfnamefont {M.}~\bibnamefont
  {Järvinen}}\ and\ \bibinfo {author} {\bibfnamefont {E.}~\bibnamefont
  {Kiritsis}},\ }\href {\doibase 10.1007/JHEP03(2012)002} {\bibfield  {journal}
  {\bibinfo  {journal} {JHEP}\ }\textbf {\bibinfo {volume} {03}},\ \bibinfo
  {pages} {002} (\bibinfo {year} {2012})},\ \Eprint
  {http://arxiv.org/abs/1112.1261} {arXiv:1112.1261 [hep-ph]} \BibitemShut
  {NoStop}%
\bibitem [{\citenamefont {Gursoy}\ and\ \citenamefont
  {Kiritsis}(2008)}]{Gursoy:2007cb}%
  \BibitemOpen
  \bibfield  {author} {\bibinfo {author} {\bibfnamefont {U.}~\bibnamefont
  {Gursoy}}\ and\ \bibinfo {author} {\bibfnamefont {E.}~\bibnamefont
  {Kiritsis}},\ }\href {\doibase 10.1088/1126-6708/2008/02/032} {\bibfield
  {journal} {\bibinfo  {journal} {JHEP}\ }\textbf {\bibinfo {volume} {02}},\
  \bibinfo {pages} {032} (\bibinfo {year} {2008})},\ \Eprint
  {http://arxiv.org/abs/0707.1324} {arXiv:0707.1324 [hep-th]} \BibitemShut
  {NoStop}%
\bibitem [{\citenamefont {Bigazzi}\ \emph {et~al.}(2005)\citenamefont
  {Bigazzi}, \citenamefont {Casero}, \citenamefont {Cotrone}, \citenamefont
  {Kiritsis},\ and\ \citenamefont {Paredes}}]{Bigazzi:2005md}%
  \BibitemOpen
  \bibfield  {author} {\bibinfo {author} {\bibfnamefont {F.}~\bibnamefont
  {Bigazzi}}, \bibinfo {author} {\bibfnamefont {R.}~\bibnamefont {Casero}},
  \bibinfo {author} {\bibfnamefont {A.~L.}\ \bibnamefont {Cotrone}}, \bibinfo
  {author} {\bibfnamefont {E.}~\bibnamefont {Kiritsis}}, \ and\ \bibinfo
  {author} {\bibfnamefont {A.}~\bibnamefont {Paredes}},\ }\href {\doibase
  10.1088/1126-6708/2005/10/012} {\bibfield  {journal} {\bibinfo  {journal}
  {JHEP}\ }\textbf {\bibinfo {volume} {10}},\ \bibinfo {pages} {012} (\bibinfo
  {year} {2005})},\ \Eprint {http://arxiv.org/abs/hep-th/0505140}
  {arXiv:hep-th/0505140 [hep-th]} \BibitemShut {NoStop}%
\bibitem [{\citenamefont {Casero}\ \emph {et~al.}(2007)\citenamefont {Casero},
  \citenamefont {Kiritsis},\ and\ \citenamefont {Paredes}}]{Casero:2007ae}%
  \BibitemOpen
  \bibfield  {author} {\bibinfo {author} {\bibfnamefont {R.}~\bibnamefont
  {Casero}}, \bibinfo {author} {\bibfnamefont {E.}~\bibnamefont {Kiritsis}}, \
  and\ \bibinfo {author} {\bibfnamefont {A.}~\bibnamefont {Paredes}},\ }\href
  {\doibase 10.1016/j.nuclphysb.2007.07.009} {\bibfield  {journal} {\bibinfo
  {journal} {Nucl. Phys.}\ }\textbf {\bibinfo {volume} {B787}},\ \bibinfo
  {pages} {98} (\bibinfo {year} {2007})},\ \Eprint
  {http://arxiv.org/abs/hep-th/0702155} {arXiv:hep-th/0702155 [HEP-TH]}
  \BibitemShut {NoStop}%
\bibitem [{\citenamefont {Alho}\ \emph {et~al.}(2013)\citenamefont {Alho},
  \citenamefont {Järvinen}, \citenamefont {Kajantie}, \citenamefont
  {Kiritsis},\ and\ \citenamefont {Tuominen}}]{Alho:2012mh}%
  \BibitemOpen
  \bibfield  {author} {\bibinfo {author} {\bibfnamefont {T.}~\bibnamefont
  {Alho}}, \bibinfo {author} {\bibfnamefont {M.}~\bibnamefont {Järvinen}},
  \bibinfo {author} {\bibfnamefont {K.}~\bibnamefont {Kajantie}}, \bibinfo
  {author} {\bibfnamefont {E.}~\bibnamefont {Kiritsis}}, \ and\ \bibinfo
  {author} {\bibfnamefont {K.}~\bibnamefont {Tuominen}},\ }\href {\doibase
  10.1007/JHEP01(2013)093} {\bibfield  {journal} {\bibinfo  {journal} {JHEP}\
  }\textbf {\bibinfo {volume} {01}},\ \bibinfo {pages} {093} (\bibinfo {year}
  {2013})},\ \Eprint {http://arxiv.org/abs/1210.4516} {arXiv:1210.4516
  [hep-ph]} \BibitemShut {NoStop}%
\bibitem [{\citenamefont {Alho}\ \emph {et~al.}(2014)\citenamefont {Alho},
  \citenamefont {Järvinen}, \citenamefont {Kajantie}, \citenamefont
  {Kiritsis}, \citenamefont {Rosen},\ and\ \citenamefont
  {Tuominen}}]{Alho:2013hsa}%
  \BibitemOpen
  \bibfield  {author} {\bibinfo {author} {\bibfnamefont {T.}~\bibnamefont
  {Alho}}, \bibinfo {author} {\bibfnamefont {M.}~\bibnamefont {Järvinen}},
  \bibinfo {author} {\bibfnamefont {K.}~\bibnamefont {Kajantie}}, \bibinfo
  {author} {\bibfnamefont {E.}~\bibnamefont {Kiritsis}}, \bibinfo {author}
  {\bibfnamefont {C.}~\bibnamefont {Rosen}}, \ and\ \bibinfo {author}
  {\bibfnamefont {K.}~\bibnamefont {Tuominen}},\ }\href {\doibase
  10.1007/JHEP02(2015)033, 10.1007/JHEP04(2014)124} {\bibfield  {journal}
  {\bibinfo  {journal} {JHEP}\ }\textbf {\bibinfo {volume} {04}},\ \bibinfo
  {pages} {124} (\bibinfo {year} {2014})},\ \bibinfo {note} {[erratum:
  JHEP02,033(2015)]},\ \Eprint {http://arxiv.org/abs/1312.5199}
  {arXiv:1312.5199 [hep-ph]} \BibitemShut {NoStop}%
\bibitem [{\citenamefont {Areán}\ \emph {et~al.}(2013)\citenamefont {Areán},
  \citenamefont {Iatrakis}, \citenamefont {Järvinen},\ and\ \citenamefont
  {Kiritsis}}]{Arean:2013tja}%
  \BibitemOpen
  \bibfield  {author} {\bibinfo {author} {\bibfnamefont {D.}~\bibnamefont
  {Areán}}, \bibinfo {author} {\bibfnamefont {I.}~\bibnamefont {Iatrakis}},
  \bibinfo {author} {\bibfnamefont {M.}~\bibnamefont {Järvinen}}, \ and\
  \bibinfo {author} {\bibfnamefont {E.}~\bibnamefont {Kiritsis}},\ }\href
  {\doibase 10.1007/JHEP11(2013)068} {\bibfield  {journal} {\bibinfo  {journal}
  {JHEP}\ }\textbf {\bibinfo {volume} {11}},\ \bibinfo {pages} {068} (\bibinfo
  {year} {2013})},\ \Eprint {http://arxiv.org/abs/1309.2286} {arXiv:1309.2286
  [hep-ph]} \BibitemShut {NoStop}%
\bibitem [{\citenamefont {Järvinen}(2015)}]{Jarvinen:2015ofa}%
  \BibitemOpen
  \bibfield  {author} {\bibinfo {author} {\bibfnamefont {M.}~\bibnamefont
  {Järvinen}},\ }\href {\doibase 10.1007/JHEP07(2015)033} {\bibfield
  {journal} {\bibinfo  {journal} {JHEP}\ }\textbf {\bibinfo {volume} {07}},\
  \bibinfo {pages} {033} (\bibinfo {year} {2015})},\ \Eprint
  {http://arxiv.org/abs/1501.07272} {arXiv:1501.07272 [hep-ph]} \BibitemShut
  {NoStop}%
\bibitem [{\citenamefont {Gursoy}\ \emph {et~al.}(2009)\citenamefont {Gursoy},
  \citenamefont {Kiritsis}, \citenamefont {Mazzanti},\ and\ \citenamefont
  {Nitti}}]{Gursoy:2009jd}%
  \BibitemOpen
  \bibfield  {author} {\bibinfo {author} {\bibfnamefont {U.}~\bibnamefont
  {Gursoy}}, \bibinfo {author} {\bibfnamefont {E.}~\bibnamefont {Kiritsis}},
  \bibinfo {author} {\bibfnamefont {L.}~\bibnamefont {Mazzanti}}, \ and\
  \bibinfo {author} {\bibfnamefont {F.}~\bibnamefont {Nitti}},\ }\href
  {\doibase 10.1016/j.nuclphysb.2009.05.017} {\bibfield  {journal} {\bibinfo
  {journal} {Nucl. Phys.}\ }\textbf {\bibinfo {volume} {B820}},\ \bibinfo
  {pages} {148} (\bibinfo {year} {2009})},\ \Eprint
  {http://arxiv.org/abs/0903.2859} {arXiv:0903.2859 [hep-th]} \BibitemShut
  {NoStop}%
\bibitem [{\citenamefont {Jokela}\ \emph {et~al.}(2019)\citenamefont {Jokela},
  \citenamefont {Järvinen},\ and\ \citenamefont {Remes}}]{Jokela:2018ers}%
  \BibitemOpen
  \bibfield  {author} {\bibinfo {author} {\bibfnamefont {N.}~\bibnamefont
  {Jokela}}, \bibinfo {author} {\bibfnamefont {M.}~\bibnamefont {Järvinen}}, \
  and\ \bibinfo {author} {\bibfnamefont {J.}~\bibnamefont {Remes}},\ }\href
  {\doibase 10.1007/JHEP03(2019)041} {\bibfield  {journal} {\bibinfo  {journal}
  {JHEP}\ }\textbf {\bibinfo {volume} {03}},\ \bibinfo {pages} {041} (\bibinfo
  {year} {2019})},\ \Eprint {http://arxiv.org/abs/1809.07770} {arXiv:1809.07770
  [hep-ph]} \BibitemShut {NoStop}%
\bibitem [{\citenamefont {Ishii}\ \emph {et~al.}(2019)\citenamefont {Ishii},
  \citenamefont {Järvinen},\ and\ \citenamefont {Nijs}}]{Ishii:2019gta}%
  \BibitemOpen
  \bibfield  {author} {\bibinfo {author} {\bibfnamefont {T.}~\bibnamefont
  {Ishii}}, \bibinfo {author} {\bibfnamefont {M.}~\bibnamefont {Järvinen}}, \
  and\ \bibinfo {author} {\bibfnamefont {G.}~\bibnamefont {Nijs}},\ }\href@noop
  {} {\  (\bibinfo {year} {2019})},\ \Eprint {http://arxiv.org/abs/1903.06169}
  {arXiv:1903.06169 [hep-ph]} \BibitemShut {NoStop}%
\bibitem [{\citenamefont {Chesler}\ \emph {et~al.}(2019)\citenamefont
  {Chesler}, \citenamefont {Jokela}, \citenamefont {Loeb},\ and\ \citenamefont
  {Vuorinen}}]{Chesler:2019osn}%
  \BibitemOpen
  \bibfield  {author} {\bibinfo {author} {\bibfnamefont {P.~M.}\ \bibnamefont
  {Chesler}}, \bibinfo {author} {\bibfnamefont {N.}~\bibnamefont {Jokela}},
  \bibinfo {author} {\bibfnamefont {A.}~\bibnamefont {Loeb}}, \ and\ \bibinfo
  {author} {\bibfnamefont {A.}~\bibnamefont {Vuorinen}},\ }\href@noop {} {\
  (\bibinfo {year} {2019})},\ \Eprint {http://arxiv.org/abs/1906.08440}
  {arXiv:1906.08440 [astro-ph.HE]} \BibitemShut {NoStop}%
\bibitem [{\citenamefont {Hoyos}\ \emph {et~al.}(2016)\citenamefont {Hoyos},
  \citenamefont {Rodríguez~Fernández}, \citenamefont {Jokela},\ and\
  \citenamefont {Vuorinen}}]{Hoyos:2016zke}%
  \BibitemOpen
  \bibfield  {author} {\bibinfo {author} {\bibfnamefont {C.}~\bibnamefont
  {Hoyos}}, \bibinfo {author} {\bibfnamefont {D.}~\bibnamefont
  {Rodríguez~Fernández}}, \bibinfo {author} {\bibfnamefont {N.}~\bibnamefont
  {Jokela}}, \ and\ \bibinfo {author} {\bibfnamefont {A.}~\bibnamefont
  {Vuorinen}},\ }\href {\doibase 10.1103/PhysRevLett.117.032501} {\bibfield
  {journal} {\bibinfo  {journal} {Phys. Rev. Lett.}\ }\textbf {\bibinfo
  {volume} {117}},\ \bibinfo {pages} {032501} (\bibinfo {year} {2016})},\
  \Eprint {http://arxiv.org/abs/1603.02943} {arXiv:1603.02943 [hep-ph]}
  \BibitemShut {NoStop}%
\bibitem [{\citenamefont {Rozali}\ \emph {et~al.}(2008)\citenamefont {Rozali},
  \citenamefont {Shieh}, \citenamefont {Van~Raamsdonk},\ and\ \citenamefont
  {Wu}}]{Rozali:2007rx}%
  \BibitemOpen
  \bibfield  {author} {\bibinfo {author} {\bibfnamefont {M.}~\bibnamefont
  {Rozali}}, \bibinfo {author} {\bibfnamefont {H.-H.}\ \bibnamefont {Shieh}},
  \bibinfo {author} {\bibfnamefont {M.}~\bibnamefont {Van~Raamsdonk}}, \ and\
  \bibinfo {author} {\bibfnamefont {J.}~\bibnamefont {Wu}},\ }\href {\doibase
  10.1088/1126-6708/2008/01/053} {\bibfield  {journal} {\bibinfo  {journal}
  {JHEP}\ }\textbf {\bibinfo {volume} {01}},\ \bibinfo {pages} {053} (\bibinfo
  {year} {2008})},\ \Eprint {http://arxiv.org/abs/0708.1322} {arXiv:0708.1322
  [hep-th]} \BibitemShut {NoStop}%
\bibitem [{\citenamefont {Li}\ \emph {et~al.}(2015)\citenamefont {Li},
  \citenamefont {Schmitt},\ and\ \citenamefont {Wang}}]{Li:2015uea}%
  \BibitemOpen
  \bibfield  {author} {\bibinfo {author} {\bibfnamefont {S.-w.}\ \bibnamefont
  {Li}}, \bibinfo {author} {\bibfnamefont {A.}~\bibnamefont {Schmitt}}, \ and\
  \bibinfo {author} {\bibfnamefont {Q.}~\bibnamefont {Wang}},\ }\href {\doibase
  10.1103/PhysRevD.92.026006} {\bibfield  {journal} {\bibinfo  {journal} {Phys.
  Rev.}\ }\textbf {\bibinfo {volume} {D92}},\ \bibinfo {pages} {026006}
  (\bibinfo {year} {2015})},\ \Eprint {http://arxiv.org/abs/1505.04886}
  {arXiv:1505.04886 [hep-ph]} \BibitemShut {NoStop}%
\bibitem [{\citenamefont {Haensel}\ and\ \citenamefont
  {Pichon}(1994)}]{Haensel:1993zw}%
  \BibitemOpen
  \bibfield  {author} {\bibinfo {author} {\bibfnamefont {P.}~\bibnamefont
  {Haensel}}\ and\ \bibinfo {author} {\bibfnamefont {B.}~\bibnamefont
  {Pichon}},\ }\href@noop {} {\bibfield  {journal} {\bibinfo  {journal}
  {Astron. Astrophys.}\ }\textbf {\bibinfo {volume} {283}},\ \bibinfo {pages}
  {313} (\bibinfo {year} {1994})},\ \Eprint
  {http://arxiv.org/abs/nucl-th/9310003} {arXiv:nucl-th/9310003 [nucl-th]}
  \BibitemShut {NoStop}%
\bibitem [{\citenamefont {Douchin}\ and\ \citenamefont
  {Haensel}(2001)}]{Douchin:2001sv}%
  \BibitemOpen
  \bibfield  {author} {\bibinfo {author} {\bibfnamefont {F.}~\bibnamefont
  {Douchin}}\ and\ \bibinfo {author} {\bibfnamefont {P.}~\bibnamefont
  {Haensel}},\ }\href {\doibase 10.1051/0004-6361:20011402} {\bibfield
  {journal} {\bibinfo  {journal} {Astron. Astrophys.}\ }\textbf {\bibinfo
  {volume} {380}},\ \bibinfo {pages} {151} (\bibinfo {year} {2001})},\ \Eprint
  {http://arxiv.org/abs/astro-ph/0111092} {arXiv:astro-ph/0111092 [astro-ph]}
  \BibitemShut {NoStop}%
\bibitem [{\citenamefont {Glendenning}(1997)}]{Glendenning:1997wn}%
  \BibitemOpen
  \bibfield  {author} {\bibinfo {author} {\bibfnamefont {N.~K.}\ \bibnamefont
  {Glendenning}},\ }\href@noop {} {\emph {\bibinfo {title} {{Compact stars:
  Nuclear physics, particle physics, and general relativity}}}}\ (\bibinfo
  {year} {1997})\BibitemShut {NoStop}%
\bibitem [{\citenamefont {Nättilä}\ \emph {et~al.}(2016)\citenamefont
  {Nättilä}, \citenamefont {Steiner}, \citenamefont {Kajava}, \citenamefont
  {Suleimanov},\ and\ \citenamefont {Poutanen}}]{Nattila:2015jra}%
  \BibitemOpen
  \bibfield  {author} {\bibinfo {author} {\bibfnamefont {J.}~\bibnamefont
  {Nättilä}}, \bibinfo {author} {\bibfnamefont {A.~W.}\ \bibnamefont
  {Steiner}}, \bibinfo {author} {\bibfnamefont {J.~J.~E.}\ \bibnamefont
  {Kajava}}, \bibinfo {author} {\bibfnamefont {V.~F.}\ \bibnamefont
  {Suleimanov}}, \ and\ \bibinfo {author} {\bibfnamefont {J.}~\bibnamefont
  {Poutanen}},\ }\href {\doibase 10.1051/0004-6361/201527416} {\bibfield
  {journal} {\bibinfo  {journal} {Astron. Astrophys.}\ }\textbf {\bibinfo
  {volume} {591}},\ \bibinfo {pages} {A25} (\bibinfo {year} {2016})},\ \Eprint
  {http://arxiv.org/abs/1509.06561} {arXiv:1509.06561 [astro-ph.HE]}
  \BibitemShut {NoStop}%
\bibitem [{\citenamefont {Nättilä}\ \emph {et~al.}(2017)\citenamefont
  {Nättilä}, \citenamefont {Miller}, \citenamefont {Steiner}, \citenamefont
  {Kajava}, \citenamefont {Suleimanov},\ and\ \citenamefont
  {Poutanen}}]{Nattila:2017wtj}%
  \BibitemOpen
  \bibfield  {author} {\bibinfo {author} {\bibfnamefont {J.}~\bibnamefont
  {Nättilä}}, \bibinfo {author} {\bibfnamefont {M.~C.}\ \bibnamefont
  {Miller}}, \bibinfo {author} {\bibfnamefont {A.~W.}\ \bibnamefont {Steiner}},
  \bibinfo {author} {\bibfnamefont {J.~J.~E.}\ \bibnamefont {Kajava}}, \bibinfo
  {author} {\bibfnamefont {V.~F.}\ \bibnamefont {Suleimanov}}, \ and\ \bibinfo
  {author} {\bibfnamefont {J.}~\bibnamefont {Poutanen}},\ }\href {\doibase
  10.1051/0004-6361/201731082} {\bibfield  {journal} {\bibinfo  {journal}
  {Astron. Astrophys.}\ }\textbf {\bibinfo {volume} {608}},\ \bibinfo {pages}
  {A31} (\bibinfo {year} {2017})},\ \Eprint {http://arxiv.org/abs/1709.09120}
  {arXiv:1709.09120 [astro-ph.HE]} \BibitemShut {NoStop}%
\bibitem [{\citenamefont {Antoniadis}\ \emph {et~al.}(2013)\citenamefont
  {Antoniadis} \emph {et~al.}}]{Antoniadis:2013pzd}%
  \BibitemOpen
  \bibfield  {author} {\bibinfo {author} {\bibfnamefont {J.}~\bibnamefont
  {Antoniadis}} \emph {et~al.},\ }\href {\doibase 10.1126/science.1233232}
  {\bibfield  {journal} {\bibinfo  {journal} {Science}\ }\textbf {\bibinfo
  {volume} {340}},\ \bibinfo {pages} {6131} (\bibinfo {year} {2013})},\ \Eprint
  {http://arxiv.org/abs/1304.6875} {arXiv:1304.6875 [astro-ph.HE]} \BibitemShut
  {NoStop}%
\bibitem [{\citenamefont {Abbott}\ \emph {et~al.}(2018)\citenamefont {Abbott}
  \emph {et~al.}}]{Abbott:2018exr}%
  \BibitemOpen
  \bibfield  {author} {\bibinfo {author} {\bibfnamefont {B.~P.}\ \bibnamefont
  {Abbott}} \emph {et~al.} (\bibinfo {collaboration} {LIGO Scientific,
  Virgo}),\ }\href {\doibase 10.1103/PhysRevLett.121.161101} {\bibfield
  {journal} {\bibinfo  {journal} {Phys. Rev. Lett.}\ }\textbf {\bibinfo
  {volume} {121}},\ \bibinfo {pages} {161101} (\bibinfo {year} {2018})},\
  \Eprint {http://arxiv.org/abs/1805.11581} {arXiv:1805.11581 [gr-qc]}
  \BibitemShut {NoStop}%
\bibitem [{\citenamefont {Bauswein}\ \emph {et~al.}(2010)\citenamefont
  {Bauswein}, \citenamefont {Janka},\ and\ \citenamefont
  {Oechslin}}]{PhysRevD.82.084043}%
  \BibitemOpen
  \bibfield  {author} {\bibinfo {author} {\bibfnamefont {A.}~\bibnamefont
  {Bauswein}}, \bibinfo {author} {\bibfnamefont {H.-T.}\ \bibnamefont {Janka}},
  \ and\ \bibinfo {author} {\bibfnamefont {R.}~\bibnamefont {Oechslin}},\
  }\href {\doibase 10.1103/PhysRevD.82.084043} {\bibfield  {journal} {\bibinfo
  {journal} {Phys. Rev. D}\ }\textbf {\bibinfo {volume} {82}},\ \bibinfo
  {pages} {084043} (\bibinfo {year} {2010})}\BibitemShut {NoStop}%
\bibitem [{\citenamefont {Gourgoulhon}\ \emph {et~al.}(2001)\citenamefont
  {Gourgoulhon}, \citenamefont {Grandclement}, \citenamefont {Taniguchi},
  \citenamefont {Marck},\ and\ \citenamefont {Bonazzola}}]{Gourgoulhon:2000nn}%
  \BibitemOpen
  \bibfield  {author} {\bibinfo {author} {\bibfnamefont {E.}~\bibnamefont
  {Gourgoulhon}}, \bibinfo {author} {\bibfnamefont {P.}~\bibnamefont
  {Grandclement}}, \bibinfo {author} {\bibfnamefont {K.}~\bibnamefont
  {Taniguchi}}, \bibinfo {author} {\bibfnamefont {J.-A.}\ \bibnamefont
  {Marck}}, \ and\ \bibinfo {author} {\bibfnamefont {S.}~\bibnamefont
  {Bonazzola}},\ }\href {\doibase 10.1103/PhysRevD.63.064029} {\bibfield
  {journal} {\bibinfo  {journal} {Phys. Rev.}\ }\textbf {\bibinfo {volume}
  {D63}},\ \bibinfo {pages} {064029} (\bibinfo {year} {2001})},\ \Eprint
  {http://arxiv.org/abs/gr-qc/0007028} {arXiv:gr-qc/0007028 [gr-qc]}
  \BibitemShut {NoStop}%
\bibitem [{\citenamefont {{Rezzolla}}\ and\ \citenamefont
  {{Zanotti}}(2013)}]{2013rehy.book.....R}%
  \BibitemOpen
  \bibfield  {author} {\bibinfo {author} {\bibfnamefont {L.}~\bibnamefont
  {{Rezzolla}}}\ and\ \bibinfo {author} {\bibfnamefont {O.}~\bibnamefont
  {{Zanotti}}},\ }\href@noop {} {\emph {\bibinfo {title} {Relativistic
  Hydrodynamics, by L.~Rezzolla and O.~Zanotti.~Oxford University Press,
  2013.~ISBN-10: 0198528906; ISBN-13: 978-0198528906}}}\ (\bibinfo {year}
  {2013})\BibitemShut {NoStop}%
\bibitem [{\citenamefont {Schnetter}\ \emph {et~al.}(2004)\citenamefont
  {Schnetter}, \citenamefont {Hawley},\ and\ \citenamefont
  {Hawke}}]{Schnetter:2003rb}%
  \BibitemOpen
  \bibfield  {author} {\bibinfo {author} {\bibfnamefont {E.}~\bibnamefont
  {Schnetter}}, \bibinfo {author} {\bibfnamefont {S.~H.}\ \bibnamefont
  {Hawley}}, \ and\ \bibinfo {author} {\bibfnamefont {I.}~\bibnamefont
  {Hawke}},\ }\href {\doibase 10.1088/0264-9381/21/6/014} {\bibfield  {journal}
  {\bibinfo  {journal} {Class. Quant. Grav.}\ }\textbf {\bibinfo {volume}
  {21}},\ \bibinfo {pages} {1465} (\bibinfo {year} {2004})},\ \Eprint
  {http://arxiv.org/abs/gr-qc/0310042} {arXiv:gr-qc/0310042 [gr-qc]}
  \BibitemShut {NoStop}%
\bibitem [{\citenamefont {Thornburg}(2004)}]{Thornburg:2003sf}%
  \BibitemOpen
  \bibfield  {author} {\bibinfo {author} {\bibfnamefont {J.}~\bibnamefont
  {Thornburg}},\ }\href {\doibase 10.1088/0264-9381/21/2/026} {\bibfield
  {journal} {\bibinfo  {journal} {Class. Quant. Grav.}\ }\textbf {\bibinfo
  {volume} {21}},\ \bibinfo {pages} {743} (\bibinfo {year} {2004})},\ \Eprint
  {http://arxiv.org/abs/gr-qc/0306056} {arXiv:gr-qc/0306056 [gr-qc]}
  \BibitemShut {NoStop}%
\bibitem [{\citenamefont {Goodale}\ \emph {et~al.}(2003)\citenamefont
  {Goodale}, \citenamefont {Allen}, \citenamefont {Lanfermann}, \citenamefont
  {Mass{\'o}}, \citenamefont {Radke}, \citenamefont {Seidel},\ and\
  \citenamefont {Shalf}}]{Goodale:2002a}%
  \BibitemOpen
  \bibfield  {author} {\bibinfo {author} {\bibfnamefont {T.}~\bibnamefont
  {Goodale}}, \bibinfo {author} {\bibfnamefont {G.}~\bibnamefont {Allen}},
  \bibinfo {author} {\bibfnamefont {G.}~\bibnamefont {Lanfermann}}, \bibinfo
  {author} {\bibfnamefont {J.}~\bibnamefont {Mass{\'o}}}, \bibinfo {author}
  {\bibfnamefont {T.}~\bibnamefont {Radke}}, \bibinfo {author} {\bibfnamefont
  {E.}~\bibnamefont {Seidel}}, \ and\ \bibinfo {author} {\bibfnamefont
  {J.}~\bibnamefont {Shalf}},\ }in\ \href {http://edoc.mpg.de/3341} {\emph
  {\bibinfo {booktitle} {Vector and Parallel Processing -- VECPAR'2002, 5th
  International Conference, Lecture Notes in Computer Science}}}\ (\bibinfo
  {publisher} {Springer},\ \bibinfo {address} {Berlin},\ \bibinfo {year}
  {2003})\BibitemShut {NoStop}%
\bibitem [{\citenamefont {Alic}\ \emph {et~al.}(2012)\citenamefont {Alic},
  \citenamefont {Bona-Casas}, \citenamefont {Bona}, \citenamefont {Rezzolla},\
  and\ \citenamefont {Palenzuela}}]{PhysRevD.85.064040}%
  \BibitemOpen
  \bibfield  {author} {\bibinfo {author} {\bibfnamefont {D.}~\bibnamefont
  {Alic}}, \bibinfo {author} {\bibfnamefont {C.}~\bibnamefont {Bona-Casas}},
  \bibinfo {author} {\bibfnamefont {C.}~\bibnamefont {Bona}}, \bibinfo {author}
  {\bibfnamefont {L.}~\bibnamefont {Rezzolla}}, \ and\ \bibinfo {author}
  {\bibfnamefont {C.}~\bibnamefont {Palenzuela}},\ }\href {\doibase
  10.1103/PhysRevD.85.064040} {\bibfield  {journal} {\bibinfo  {journal} {Phys.
  Rev. D}\ }\textbf {\bibinfo {volume} {85}},\ \bibinfo {pages} {064040}
  (\bibinfo {year} {2012})}\BibitemShut {NoStop}%
\bibitem [{\citenamefont {Brown}\ \emph {et~al.}(2009)\citenamefont {Brown},
  \citenamefont {Diener}, \citenamefont {Sarbach}, \citenamefont {Schnetter},\
  and\ \citenamefont {Tiglio}}]{PhysRevD.79.044023}%
  \BibitemOpen
  \bibfield  {author} {\bibinfo {author} {\bibfnamefont {D.}~\bibnamefont
  {Brown}}, \bibinfo {author} {\bibfnamefont {P.}~\bibnamefont {Diener}},
  \bibinfo {author} {\bibfnamefont {O.}~\bibnamefont {Sarbach}}, \bibinfo
  {author} {\bibfnamefont {E.}~\bibnamefont {Schnetter}}, \ and\ \bibinfo
  {author} {\bibfnamefont {M.}~\bibnamefont {Tiglio}},\ }\href {\doibase
  10.1103/PhysRevD.79.044023} {\bibfield  {journal} {\bibinfo  {journal} {Phys.
  Rev. D}\ }\textbf {\bibinfo {volume} {79}},\ \bibinfo {pages} {044023}
  (\bibinfo {year} {2009})}\BibitemShut {NoStop}%
\bibitem [{\citenamefont {Löffler}\ \emph {et~al.}(2012)\citenamefont
  {Löffler}, \citenamefont {Faber}, \citenamefont {Bentivegna}, \citenamefont
  {Bode}, \citenamefont {Diener}, \citenamefont {Haas}, \citenamefont {Hinder},
  \citenamefont {Mundim}, \citenamefont {Ott}, \citenamefont {Schnetter},
  \citenamefont {Allen}, \citenamefont {Campanelli},\ and\ \citenamefont
  {Laguna}}]{L_ffler_2012}%
  \BibitemOpen
  \bibfield  {author} {\bibinfo {author} {\bibfnamefont {F.}~\bibnamefont
  {Löffler}}, \bibinfo {author} {\bibfnamefont {J.}~\bibnamefont {Faber}},
  \bibinfo {author} {\bibfnamefont {E.}~\bibnamefont {Bentivegna}}, \bibinfo
  {author} {\bibfnamefont {T.}~\bibnamefont {Bode}}, \bibinfo {author}
  {\bibfnamefont {P.}~\bibnamefont {Diener}}, \bibinfo {author} {\bibfnamefont
  {R.}~\bibnamefont {Haas}}, \bibinfo {author} {\bibfnamefont {I.}~\bibnamefont
  {Hinder}}, \bibinfo {author} {\bibfnamefont {B.~C.}\ \bibnamefont {Mundim}},
  \bibinfo {author} {\bibfnamefont {C.~D.}\ \bibnamefont {Ott}}, \bibinfo
  {author} {\bibfnamefont {E.}~\bibnamefont {Schnetter}}, \bibinfo {author}
  {\bibfnamefont {G.}~\bibnamefont {Allen}}, \bibinfo {author} {\bibfnamefont
  {M.}~\bibnamefont {Campanelli}}, \ and\ \bibinfo {author} {\bibfnamefont
  {P.}~\bibnamefont {Laguna}},\ }\href {\doibase
  10.1088/0264-9381/29/11/115001} {\bibfield  {journal} {\bibinfo  {journal}
  {Classical and Quantum Gravity}\ }\textbf {\bibinfo {volume} {29}},\ \bibinfo
  {pages} {115001} (\bibinfo {year} {2012})}\BibitemShut {NoStop}%
\bibitem [{\citenamefont {Alcubierre}\ \emph {et~al.}(2003)\citenamefont
  {Alcubierre}, \citenamefont {Br\"ugmann}, \citenamefont {Diener},
  \citenamefont {Koppitz}, \citenamefont {Pollney}, \citenamefont {Seidel},\
  and\ \citenamefont {Takahashi}}]{PhysRevD.67.084023}%
  \BibitemOpen
  \bibfield  {author} {\bibinfo {author} {\bibfnamefont {M.}~\bibnamefont
  {Alcubierre}}, \bibinfo {author} {\bibfnamefont {B.}~\bibnamefont
  {Br\"ugmann}}, \bibinfo {author} {\bibfnamefont {P.}~\bibnamefont {Diener}},
  \bibinfo {author} {\bibfnamefont {M.}~\bibnamefont {Koppitz}}, \bibinfo
  {author} {\bibfnamefont {D.}~\bibnamefont {Pollney}}, \bibinfo {author}
  {\bibfnamefont {E.}~\bibnamefont {Seidel}}, \ and\ \bibinfo {author}
  {\bibfnamefont {R.}~\bibnamefont {Takahashi}},\ }\href {\doibase
  10.1103/PhysRevD.67.084023} {\bibfield  {journal} {\bibinfo  {journal} {Phys.
  Rev. D}\ }\textbf {\bibinfo {volume} {67}},\ \bibinfo {pages} {084023}
  (\bibinfo {year} {2003})}\BibitemShut {NoStop}%
\bibitem [{\citenamefont {Pollney}\ \emph {et~al.}(2007)\citenamefont
  {Pollney}, \citenamefont {Reisswig}, \citenamefont {Rezzolla}, \citenamefont
  {Szil\'agyi}, \citenamefont {Ansorg}, \citenamefont {Deris}, \citenamefont
  {Diener}, \citenamefont {Dorband}, \citenamefont {Koppitz}, \citenamefont
  {Nagar},\ and\ \citenamefont {Schnetter}}]{PhysRevD.76.124002}%
  \BibitemOpen
  \bibfield  {author} {\bibinfo {author} {\bibfnamefont {D.}~\bibnamefont
  {Pollney}}, \bibinfo {author} {\bibfnamefont {C.}~\bibnamefont {Reisswig}},
  \bibinfo {author} {\bibfnamefont {L.}~\bibnamefont {Rezzolla}}, \bibinfo
  {author} {\bibfnamefont {B.}~\bibnamefont {Szil\'agyi}}, \bibinfo {author}
  {\bibfnamefont {M.}~\bibnamefont {Ansorg}}, \bibinfo {author} {\bibfnamefont
  {B.}~\bibnamefont {Deris}}, \bibinfo {author} {\bibfnamefont
  {P.}~\bibnamefont {Diener}}, \bibinfo {author} {\bibfnamefont {E.~N.}\
  \bibnamefont {Dorband}}, \bibinfo {author} {\bibfnamefont {M.}~\bibnamefont
  {Koppitz}}, \bibinfo {author} {\bibfnamefont {A.}~\bibnamefont {Nagar}}, \
  and\ \bibinfo {author} {\bibfnamefont {E.}~\bibnamefont {Schnetter}},\ }\href
  {\doibase 10.1103/PhysRevD.76.124002} {\bibfield  {journal} {\bibinfo
  {journal} {Phys. Rev. D}\ }\textbf {\bibinfo {volume} {76}},\ \bibinfo
  {pages} {124002} (\bibinfo {year} {2007})}\BibitemShut {NoStop}%
\bibitem [{\citenamefont {Banyuls}\ \emph {et~al.}(1997)\citenamefont
  {Banyuls}, \citenamefont {Font}, \citenamefont {Ibanez}, \citenamefont
  {Marti},\ and\ \citenamefont {Miralles}}]{Banyuls_1997}%
  \BibitemOpen
  \bibfield  {author} {\bibinfo {author} {\bibfnamefont {F.}~\bibnamefont
  {Banyuls}}, \bibinfo {author} {\bibfnamefont {J.~A.}\ \bibnamefont {Font}},
  \bibinfo {author} {\bibfnamefont {J.~M.}\ \bibnamefont {Ibanez}}, \bibinfo
  {author} {\bibfnamefont {J.~M.}\ \bibnamefont {Marti}}, \ and\ \bibinfo
  {author} {\bibfnamefont {J.~A.}\ \bibnamefont {Miralles}},\ }\href {\doibase
  10.1086/303604} {\bibfield  {journal} {\bibinfo  {journal} {The Astrophysical
  Journal}\ }\textbf {\bibinfo {volume} {476}},\ \bibinfo {pages} {221}
  (\bibinfo {year} {1997})}\BibitemShut {NoStop}%
\bibitem [{\citenamefont {Radice}\ and\ \citenamefont
  {Rezzolla}(2012)}]{Radice:2012cu}%
  \BibitemOpen
  \bibfield  {author} {\bibinfo {author} {\bibfnamefont {D.}~\bibnamefont
  {Radice}}\ and\ \bibinfo {author} {\bibfnamefont {L.}~\bibnamefont
  {Rezzolla}},\ }\href {\doibase 10.1051/0004-6361/201219735} {\bibfield
  {journal} {\bibinfo  {journal} {Astron. Astrophys.}\ }\textbf {\bibinfo
  {volume} {547}},\ \bibinfo {pages} {A26} (\bibinfo {year} {2012})},\ \Eprint
  {http://arxiv.org/abs/1206.6502} {arXiv:1206.6502 [astro-ph.IM]} \BibitemShut
  {NoStop}%
\bibitem [{\citenamefont {Radice}\ \emph
  {et~al.}(2014{\natexlab{a}})\citenamefont {Radice}, \citenamefont
  {Rezzolla},\ and\ \citenamefont {Galeazzi}}]{Radice:2013hxh}%
  \BibitemOpen
  \bibfield  {author} {\bibinfo {author} {\bibfnamefont {D.}~\bibnamefont
  {Radice}}, \bibinfo {author} {\bibfnamefont {L.}~\bibnamefont {Rezzolla}}, \
  and\ \bibinfo {author} {\bibfnamefont {F.}~\bibnamefont {Galeazzi}},\ }\href
  {\doibase 10.1093/mnrasl/slt137} {\bibfield  {journal} {\bibinfo  {journal}
  {Mon. Not. Roy. Astron. Soc.}\ }\textbf {\bibinfo {volume} {437}},\ \bibinfo
  {pages} {L46} (\bibinfo {year} {2014}{\natexlab{a}})},\ \Eprint
  {http://arxiv.org/abs/1306.6052} {arXiv:1306.6052 [gr-qc]} \BibitemShut
  {NoStop}%
\bibitem [{\citenamefont {Radice}\ \emph
  {et~al.}(2014{\natexlab{b}})\citenamefont {Radice}, \citenamefont
  {Rezzolla},\ and\ \citenamefont {Galeazzi}}]{Radice:2013xpa}%
  \BibitemOpen
  \bibfield  {author} {\bibinfo {author} {\bibfnamefont {D.}~\bibnamefont
  {Radice}}, \bibinfo {author} {\bibfnamefont {L.}~\bibnamefont {Rezzolla}}, \
  and\ \bibinfo {author} {\bibfnamefont {F.}~\bibnamefont {Galeazzi}},\ }\href
  {\doibase 10.1088/0264-9381/31/7/075012} {\bibfield  {journal} {\bibinfo
  {journal} {Class. Quant. Grav.}\ }\textbf {\bibinfo {volume} {31}},\ \bibinfo
  {pages} {075012} (\bibinfo {year} {2014}{\natexlab{b}})},\ \Eprint
  {http://arxiv.org/abs/1312.5004} {arXiv:1312.5004 [gr-qc]} \BibitemShut
  {NoStop}%
\bibitem [{\citenamefont {Radice}\ \emph {et~al.}(2015)\citenamefont {Radice},
  \citenamefont {Rezzolla},\ and\ \citenamefont {Galeazzi}}]{Radice:2015nva}%
  \BibitemOpen
  \bibfield  {author} {\bibinfo {author} {\bibfnamefont {D.}~\bibnamefont
  {Radice}}, \bibinfo {author} {\bibfnamefont {L.}~\bibnamefont {Rezzolla}}, \
  and\ \bibinfo {author} {\bibfnamefont {F.}~\bibnamefont {Galeazzi}},\
  }\bibfield  {booktitle} {\emph {\bibinfo {booktitle} {{Proceedings, Numerical
  Modeling of Space Plasma Flows (ASTRONUM-2014): Long Beach, CA, USA, June
  23-27, 2014}}},\ }\href@noop {} {\bibfield  {journal} {\bibinfo  {journal}
  {ASP Conf. Ser.}\ }\textbf {\bibinfo {volume} {498}},\ \bibinfo {pages} {121}
  (\bibinfo {year} {2015})},\ \Eprint {http://arxiv.org/abs/1502.00551}
  {arXiv:1502.00551 [gr-qc]} \BibitemShut {NoStop}%
\bibitem [{\citenamefont {{Radice, D.}}\ and\ \citenamefont {{Rezzolla,
  L.}}(2012)}]{refId0}%
  \BibitemOpen
  \bibfield  {author} {\bibinfo {author} {\bibnamefont {{Radice, D.}}}\ and\
  \bibinfo {author} {\bibnamefont {{Rezzolla, L.}}},\ }\href {\doibase
  10.1051/0004-6361/201219735} {\bibfield  {journal} {\bibinfo  {journal}
  {A\&A}\ }\textbf {\bibinfo {volume} {547}},\ \bibinfo {pages} {A26} (\bibinfo
  {year} {2012})}\BibitemShut {NoStop}%
\bibitem [{\citenamefont {Press}\ \emph {et~al.}(2007)\citenamefont {Press},
  \citenamefont {Teukolsky}, \citenamefont {Vetterling},\ and\ \citenamefont
  {Flannery}}]{Press:2007:NRE:1403886}%
  \BibitemOpen
  \bibfield  {author} {\bibinfo {author} {\bibfnamefont {W.~H.}\ \bibnamefont
  {Press}}, \bibinfo {author} {\bibfnamefont {S.~A.}\ \bibnamefont
  {Teukolsky}}, \bibinfo {author} {\bibfnamefont {W.~T.}\ \bibnamefont
  {Vetterling}}, \ and\ \bibinfo {author} {\bibfnamefont {B.~P.}\ \bibnamefont
  {Flannery}},\ }\href@noop {} {\emph {\bibinfo {title} {Numerical Recipes 3rd
  Edition: The Art of Scientific Computing}}},\ \bibinfo {edition} {3rd}\ ed.\
  (\bibinfo  {publisher} {Cambridge University Press},\ \bibinfo {address} {New
  York, NY, USA},\ \bibinfo {year} {2007})\BibitemShut {NoStop}%
\bibitem [{\citenamefont {Alcubierre}(2006)}]{article}%
  \BibitemOpen
  \bibfield  {author} {\bibinfo {author} {\bibfnamefont {M.}~\bibnamefont
  {Alcubierre}},\ }\href {\doibase 10.1093/acprof:oso/9780199205677.001.0001}
  {\bibfield  {journal} {\bibinfo  {journal} {Introduction to 3+1 Numerical
  Relativity}\ } (\bibinfo {year} {2006}),\
  10.1093/acprof:oso/9780199205677.001.0001}\BibitemShut {NoStop}%
\bibitem [{Note1()}]{Note1}%
  \BibitemOpen
  \bibinfo {note} {Movies of the mergers can be found at \protect \href
  {https://sites.google.com/site/wilkevanderschee/neutron-stars}{sites.google.com/site/wilkevanderschee/neutron-stars}}\BibitemShut
  {NoStop}%
\bibitem [{\citenamefont {Barsotti}\ \emph {et~al.}(2018)\citenamefont
  {Barsotti}, \citenamefont {Gras}, \citenamefont {Evans},\ and\ \citenamefont
  {Fritschel}}]{adLIGO2018}%
  \BibitemOpen
  \bibfield  {author} {\bibinfo {author} {\bibfnamefont {L.}~\bibnamefont
  {Barsotti}}, \bibinfo {author} {\bibfnamefont {S.}~\bibnamefont {Gras}},
  \bibinfo {author} {\bibfnamefont {M.}~\bibnamefont {Evans}}, \ and\ \bibinfo
  {author} {\bibfnamefont {P.}~\bibnamefont {Fritschel}},\ }\href@noop {}
  {\bibfield  {journal} {\bibinfo  {journal} {LIGO Document No. T1800044-v5}\ }
  (\bibinfo {year} {2018})}\BibitemShut {NoStop}%
\bibitem [{\citenamefont {Punturo}\ \emph {et~al.}(2010)\citenamefont {Punturo}
  \emph {et~al.}}]{Punturo:2010zza}%
  \BibitemOpen
  \bibfield  {author} {\bibinfo {author} {\bibfnamefont {M.}~\bibnamefont
  {Punturo}} \emph {et~al.},\ }\bibfield  {booktitle} {\emph {\bibinfo
  {booktitle} {{Gravitational waves. Proceedings, 8th Edoardo Amaldi
  Conference, Amaldi 8, New York, USA, June 22-26, 2009}}},\ }\href {\doibase
  10.1088/0264-9381/27/8/084007} {\bibfield  {journal} {\bibinfo  {journal}
  {Class. Quant. Grav.}\ }\textbf {\bibinfo {volume} {27}},\ \bibinfo {pages}
  {084007} (\bibinfo {year} {2010})}\BibitemShut {NoStop}%
\end{thebibliography}%

\end{document}